\newif\iffigs\figstrue

\documentclass[12pt]{article}
\iffigs
  \input epsf
\else
\fi

\batchmode
\newfont{\footscrfont}{rsfs10}
  \newfont{\footbbbfont}{msbm10}
  \newfont{\manfont}{manfnt}
\errorstopmode

\newif\ifscrf\scrftrue
\ifx\footscrfont\nullfont
  \scrffalse
\fi

\newif\ifamsf\amsftrue
\ifx\footbbbfont\nullfont
  \amsffalse
\fi

\def\ppnumber{\vbox{\baselineskip14pt\hbox{YITP-SB 00-76}
\hbox{hep-th/0011257}}}
\def\ppdate{November 2000}
\def\pplogo{\vbox{\kern-\headheight\kern -15pt
\halign{##&##\hfil\cr&{
\ppnumber}\cr\rule{0pt}{2.5ex}&\ppdate\cr}
}}

\makeatletter
\date{}
\def\dedicatory#1{\def\@date{\normalsize\it#1}}
\def\subjclass#1{\def\@thefnmark{}\@footnotetext{1991
    {\it Mathematics Subject Classification.} #1}}
\def\keywords#1{\def\@thefnmark{}\@footnotetext{
    {\it Key words and phrases.} #1}}

\def\ps@firstpage{\ps@empty \def\@oddhead{\hss\pplogo}%
  \let\@evenhead\@oddhead 
}
\def\maketitle{\par
 \begingroup
 \def\thefootnote{\fnsymbol{footnote}}
 \def\@makefnmark{\hbox
 to 0pt{$^{\@thefnmark}$\hss}}
 \if@twocolumn
 \twocolumn[\@maketitle]
 \else \newpage
 \global\@topnum\z@ \@maketitle \fi\thispagestyle{firstpage}\@thanks
 \endgroup
 \setcounter{footnote}{0}
 \let\maketitle\relax
 \let\@maketitle\relax
 \gdef\@thanks{}\gdef\@author{}\gdef\@title{}\let\thanks\relax}

\def\abstract{\if@twocolumn
\section*{Abstract}
\else \small
\begin{center}
{\bf ABSTRACT}
\end{center}
\quotation
\fi}

\def\thebibliography#1{\section*{References\@mkboth
 {REFERENCES}{REFERENCES}}\small\list
 {[\arabic{enumi}]}{\settowidth\labelwidth{[#1]}\leftmargin\labelwidth
 \advance\leftmargin\labelsep
 \usecounter{enumi}}
 \def\newblock{\hskip .11em plus .33em minus .07em}
 \sloppy\clubpenalty4000\widowpenalty4000
 \sfcode`\.=1000\relax}

\newif\iffn\fnfalse

\@ifundefined{reset@font}{\let\reset@font\empty}{} 
\long\def\@footnotetext#1{\insert\footins{\reset@font\footnotesize
    \interlinepenalty\interfootnotelinepenalty
    \splittopskip\footnotesep
    \splitmaxdepth \dp\strutbox \floatingpenalty \@MM
    \hsize\columnwidth \@parboxrestore
   \edef\@currentlabel{\csname p@footnote\endcsname\@thefnmark}\@makefntext
    {\rule{\z@}{\footnotesep}\ignorespaces
      \fntrue#1\fnfalse\strut}}}

\makeatother

\ifamsf
  \newfont{\bigbbbfont}{msbm10 scaled\magstep2}
  \newfont{\bbbfont}{msbm10 scaled\magstep1}  
  \newfont{\smallbbbfont}{msbm8}
  \newfont{\tinybbbfont}{msbm6}
  \newfont{\smallfootbbbfont}{msbm7}
  \newfont{\tinyfootbbbfont}{msbm5}
  \newfont{\biggthfont}{eufm10 scaled\magstep2}
  \newfont{\gthfont}{eufm10 scaled\magstep1}  
  \newfont{\smallgthfont}{eufm8}
  \newfont{\tinygthfont}{eufm6}
  \newfont{\footgthfont}{eufm10}
  \newfont{\smallfootgthfont}{eufm7}
  \newfont{\tinyfootgthfont}{eufm5}
\fi

\ifscrf
  \newfont{\scrfont}{rsfs10 scaled\magstep1}  
  \newfont{\smallscrfont}{rsfs7}
  \newfont{\tinyscrfont}{rsfs7}
  \newfont{\smallfootscrfont}{rsfs7}
  \newfont{\tinyfootscrfont}{rsfs7}
\fi

\ifamsf
  \newcommand{\Bbb}[1]{\iffn
      \mathchoice{\mbox{\footbbbfont #1}}{\mbox{\footbbbfont #1}}
      {\mbox{\smallfootbbbfont #1}}{\mbox{\tinyfootbbbfont #1}}\else
      \mathchoice{\mbox{\bbbfont #1}}{\mbox{\bbbfont #1}}
      {\mbox{\smallbbbfont #1}}{\mbox{\tinybbbfont #1}}\fi}
  
\else
  \def\bigbbbfont{\bf}
  \def\Bbb{\bf}
  
\fi

\ifscrf
  \newcommand{\Scr}[1]{\iffn
    \mathchoice{\mbox{\footscrfont #1}}{\mbox{\footscrfont #1}}
    {\mbox{\smallfootscrfont #1}}{\mbox{\tinyfootscrfont #1}}\else
    \mathchoice{\mbox{\scrfont #1}}{\mbox{\scrfont #1}}
    {\mbox{\smallscrfont #1}}{\mbox{\tinyscrfont #1}}\fi}
\else
  \def\Scr{\cal}
\fi

\def\C{{\Bbb C}}

\def\R{{\Bbb R}}
\def\Z{{\Bbb Z}}

\def\bearray{\begin{eqnarray}}
\def\eearray{\end{eqnarray}}
\def\bearraynn{\begin{eqnarray*}}
\def\eearraynn{\end{eqnarray*}}
\def\bfig{\begin{figure}}
\def\efig{\end{figure}}

\def\opeq#1{\advance\lineskip#1 \advance\baselineskip#1
        \advance\lineskiplimit#1}

\def\cM{{\Scr M}}

\def\cD{{\Scr D}}

\def\cMc{{\hfuzz=100cm\hbox to 0pt{$\;\overline{\phantom{X}}$}\cM}}
\def\barcD{{\hfuzz=100cm\hbox to 0pt{$\;\overline{\phantom{X}}$}\cD}}

\ifamsf

\else

\fi

\def\boldone{\relax{\rm 1\kern-.35em 1}}

\newtheorem{Proposition}{Proposition}[section]

\newtheorem{Theorem}{Theorem}[section]
\newtheorem{Lemma}{Lemma}[section]
\newtheorem{Corrolary}{Corrolary}[section]

\newcommand{\be}{\begin{equation}}
\newcommand{\ee}{\end{equation}}
\newcommand{\bea}{\begin{eqnarray}}
\newcommand{\eea}{\end{eqnarray}}

\newcommand{\bp}{\begin{Proposition}}
\newcommand{\ep}{\end{Proposition}}
\newcommand{\bt}{\begin{Theorem}}
\newcommand{\et}{\end{Theorem}}
\newcommand{\bl}{\begin{Lemma}}
\newcommand{\el}{\end{Lemma}}
\newcommand{\bc}{\begin{Corrolary}}
\newcommand{\ec}{\end{Corrolary}}
\newcommand{\nn}{\nonumber}

\usepackage{graphics}

\begin{document}

\title{Instanton amplitudes in open-closed topological string theory}

\author{C.~I.~Lazaroiu$^{*}$}

\date{}

\maketitle

\vbox{ \centerline{C.~N.~Yang Institute for Theoretical Physics} 
\centerline{SUNY at Stony Brook} 
\centerline{NY11794-3840, U.S.A.} 
\medskip 
\medskip
\bigskip }

\abstract{I use the universal instanton formalism to 
discuss quantum effects in the open-closed topological string 
theory of a Calabi-Yau A-model,  in the presence of a multiply-wrapped
`Floer' D-brane. This gives a precise meaning (up to the issue of compactifying 
the relevant moduli spaces) to the instanton corrections
which affect sigma model and topological string amplitudes. The cohomological 
formalism I use recovers the homological 
approach used by Fukaya and collaborators in the singly-wrapped case, even though it is not a 
naive 
generalization of the latter.
I also prove some 
non-renormalization theorems for amplitudes with low number of insertions.
The non-renormalization argument is purely geometric 
and based on the universal instanton formulation,
and thus it does not assume that the background satisfies the string equations of 
motion. These 
results are valid even though the D-brane background typically 
receives worldsheet instanton corrections. I also point out that the localized form of the 
boundary BRST operator receives instanton corrections and make a few comments on the 
consequences of this effect.}

\vskip .6in

$*$ calin@insti.physics.sunysb.edu

\pagebreak

\tableofcontents

\pagebreak

\section{Introduction}

Topological sigma  models in the  presence of D-branes form  an active
area of research due to  their relevance for open string extensions of
mirror  symmetry \cite{Vafa, Kontsevich,Kontsevich_recent}  
and  in  general  for the  physics  of  D-branes  in
Calabi-Yau compactifications \cite{Douglas_quintic,Douglas_Kontsevich}.  
Much recent work in this  direction is based
on the  paper \cite{Witten_CS}, where  boundary conditions for
such models  were first  considered. Unfortunately, recent studies of
the subject suffer from a lack of systematic
development  of  the  basic framework of topological open-closed strings. 

The purpose of the present paper is to carry out some of the required analysis  
for the open-closed topological  $A$ model  and its associated string theory
in the presence  of a (multiply-wrapped) `Floer' D-brane. This is 
a preliminary step which is necessary before one can 
consistently  formulate the string field theory of open-closed 
topological A strings. A more detailed analysis of this theory is deferred 
to a companion paper \cite{sft}.

Before discussing string theory  issues, one has to consider the more basic
problem of localization for topological  sigma model and string amplitudes in
the presence of disk instanton corrections.  
A clear formulation of localization does not seem to have been given
for the multiply wrapped case
\footnote{The definition
of the `number of disk  instantons' is slightly non-obvious 
in this situation, and even the  singly-wrapped case 
requires some careful analysis.  Since 
the string field theory action  of \cite{Witten_CS} does not take into
account  all of  the relevant  instanton effects,
formal expressions obtained  by differentiating the partition 
function of \cite{Witten_CS} as in \cite{Vafa_mirror, Kachru}
can be misleading. This is true even for the case of D-branes wrapped once 
around 
{\em special} Lagrangian cycles, due to the existence of further 
instanton corrections 
which are responsible for the obstructions noticed in \cite{Fukaya, Kachru, Fukaya2}. 
In the work of \cite{Vafa, Kachru}, one is faced with the situation of 
recovering 
disk instanton amplitudes from
a string field theory action which is in contradiction with 
the existence of these obstructions. 
Clearly one needs an independent analysis of instanton amplitudes 
and a construction of the string field 
theory which takes all 
such effects into account. This can be achieved by 
building the string field theory around a background which does 
not satisfy the equations of motion, as I shall discuss in more detail in 
\cite{sft}.}. 
One of the aims of the present paper is to approach this problem
from a `Lagrangian'  point of view (similar to the one
adopted  in  \cite{Witten_NLSM,Witten_mirror}),   as  opposed  to  the
possibly more direct, but  considerably less explicit Hamiltonian approach 
followed
originally in \cite{Witten_CS}. This enables us to give a clear 
description 
of what the instanton amplitudes compute in the presence of a bundle on 
the D-brane's worldvolume, and has the added advantage that it does not 
assume that the open string background satisfies the string equations of 
motion. As I discuss in more detail in \cite{sft}, these results can be 
used to recover the correct description of on-shell observables in the 
presence of  instanton corrections. 

A precise formulation of instanton corrections to open string amplitudes 
was proposed in \cite{Fukaya, Fukaya2}. Unfortunately, the homological 
approach of \cite{Fukaya} seems to be limited to the singly-wrapped 
case, where Poincare duality can be used in its traditional sense.
A naive extension
of this approach is problematic for multiply wrapped
D-branes, since Poincare duality cannot be used directly for bundle-valued forms.
Instead, I will retreat to a cohomological formulation in terms 
of cup products of bundle valued forms defined over the instanton moduli space, which 
also has the advantage that it is more directly related to the physical 
interpretation of amplitudes.
This  requires a careful discussion of
localization and  the use of the universal  instanton formalism, which
is   carried  out for the open-closed sigma model in   Section 3, 
and for the associated topological string theory in Section 4.  
The result reduces to that of \cite{Fukaya, Fukaya2} 
for the particular case of singly-wrapped D-branes.

As noticed in \cite{Fukaya} and rediscovered in \cite{Douglas_quintic,Kachru}, 
the associated open string background will typically receive quantum corrections.
This means that one is building the string field theory around a 
background which fails to satisfy the instanton-corrected string 
equations of motion. In particular, the BRST operator fails to localize on 
its large radius representative, which implies that the BRST cohomology of 
the model as computed in \cite{Witten_CS} has to be modified. 
The relevant corrections and deformation 
theory will be discussed in more detail in \cite{sft} from a string field 
theoretic perspective, where in particular I will show that the 
obstruction of \cite{Fukaya, Fukaya2} has a very simple string-field theoretic 
interpretation. Because the analysis of \cite{sft} involves a string 
field theory expanded around the wrong vacuum, one must be careful not to use
the final result when constructing the theory. In particular, it is important to 
correctly identify the boundary topological metric and give an argument 
for its independence of Calabi-Yau radius which does not rely on BRST closure 
and conformal 
invariance\footnote{This is to say that we cannot apply 
the open string version of the argument of \cite{DVV}, since this 
assumes that we have correctly identified a conformally invariant 
string vacuum and the associated BRST charge. Since we only know 
how to recover this data from string field theory arguments, this 
would amount to using the conclusion as a hypothesis. In particular, we cannot 
simply borrow the results of \cite{Hofman}.}. To avoid circular 
arguments, we give a non-renormalization theorem which is based on 
the geometry of the associated moduli spaces. This result
assures that the boundary topological metric does not receive instanton 
corrections, a statement which is important for building the open string field 
theory \cite{sft}. 
I also give non-renormalization results for the bulk and boundary 
one-point amplitudes
on the disk. Finally, I discuss the relation between string amplitudes and 
nonlinear sigma model amplitudes with integrated insertions of descendants. 
Using explicit expressions after localization allows for
a proof of their equivalence which does not assume that the background 
satisfies the string equations of motion. Section 2 reviews the structure 
of the open-closed A model. Appendix A re-derives localization  for the BRST operator 
in the large radius limit in the Lagrangian framework, while 
Appendix B discusses the Maslow index 
of disk instantons. 

A word of caution is in order for the mathematically inclined reader. 
In this paper, I 
treat the various instanton moduli spaces naively, i.e. I neglect 
the fact that many of the arguments I use are not rigorous unless one 
compactifies these spaces appropriately. 
The mathematical machinery needed for approaching this problem 
is extremely complex and has recently become available in the book 
\cite{Fukaya2}, to which I refer 
the interested reader. To a large extent, the present work and its 
sequel \cite{sft} are an attempt to fill in the holes separating the
work of the physics community from that of \cite{Fukaya} and \cite{Fukaya2}.

\section{The open topological A-model around a classical background}

\subsection{Review of the model in the presence of one D-brane}

Consider a 
complex d-dimensional Calabi-Yau manifold $X$, whose complexified tangent bundle 
we denote by ${\cal T}X=TX\oplus {\overline T}X$, where $TX$ and 
${\overline T}X$ are the holomorphic and antiholomorphic tangent bundles.
Recall that the nonlinear sigma model associated to $X$ contains 
worldsheet bosons described by a map $\phi$ from the worldsheet to $X$ 
as well as  fermions 
$\psi_{L},\psi_{R}$, which are sections of the bundles 
$K_{c}^{1/2}\otimes\phi^*({\cal T}X)=
K_{c}^{1/2}\otimes\phi^*(TX)\oplus K_{c}^{1/2}\otimes 
\phi^*({\overline T}X)$ and 
$K_{a}^{1/2}\otimes \phi^*({\cal T}X)=
K_{a}^{1/2}\otimes \phi^*(TX)\oplus K_{a}^{1/2}\otimes 
\phi^*({\overline T}X)$. 
Here $K_c,K_a$ are the canonical and anticanonical line bundles 
on the worldsheet while $K_{c}^{1/2}$ and $K_{a}^{1/2}$ are 
choices for their square rots, which control the spin structure. 

To build the associated  A-model, one simply declares these fields to be 
sections of the `twisted' bundles  $\phi^*(TX)\oplus 
K_{c}\otimes \phi^*({\overline T}X)$ and $K_a\otimes \phi^*(TX)\oplus \phi^*
({\overline T}X)$ \cite{Witten_mirror}. 
The twisted fields (which we denote by the same 
symbols) admit decompositions:   
\bea
\label{decomp}
\psi_{L}&=&\chi_{L}+\lambda_{L}\\
\psi_{R}&=&\chi_{R}+\lambda_{R}~~,
\eea
where $\chi_L$ and $\chi_R$ are 
sections of $\phi^*(TX)$ and $\phi^*({\overline T}X)$ 
and $\lambda_L,\lambda_R$ are 
sections of  $K_c\otimes \phi^*({\overline T}X)$ and $K_a\otimes \phi^*(TX)$.

The bulk action of the twisted model (on the disk $D$) is given by:
{\scriptsize\bea
S=\int_{D}{d^2z\left[\frac{1}{2}g(\partial_z\phi(z,{\overline z}), 
\partial_{\overline z}\phi(z,{\overline z}))
+ig(\lambda_L(z,{\overline z}),\partial_{\overline z}\chi_L(z,{\overline z}))+
ig(\lambda_R(z,{\overline z}),\partial_z\chi_R(z,{\overline z}))\right]}+\nn\\
+\int_{D}{d^2z R(\chi_L(z,{\overline z}),
\lambda_L(z,{\overline z}),\lambda_R(z,{\overline z}),
\chi_R(z,{\overline z}))}~~,
\eea}\noindent where we wrote $\lambda_L=\lambda_L(z,{\overline z})d_z$, 
$\lambda_R=\lambda_R(z,{\overline z})d{\overline z}$ etc. Note that 
$\lambda_L$ etc need not be holomorphic sections of the corresponding 
bundles, since we are not imposing the equations of motion.

For later use, we also define:
\bea
\chi=\chi_L+\chi_R~~\\
\lambda=\lambda_L+\lambda_R~~,
\eea
which are sections of  $\phi^*({\cal  T}X)$ and  $K_{c}\otimes\phi^*(
{\overline T}X)\oplus K_{a}\otimes \phi^*(TX)$, respectively.

The A model admits a nilpotent 
BRST symmetry, whose action on the fields is given by:
\bea
\delta_{Q}~\phi&=&i\xi\chi~~\\
\delta_Q\chi~~&=&0~~\\
\delta_{Q} \lambda_{L}^{\overline i}&=&-\xi \partial 
\phi^{\overline i}-i\xi\chi^{\overline j}
\Gamma^{\overline i}_{{\overline j}{\overline m}}\lambda_{L}^{\overline m}~~\\
\delta_{Q} \lambda_{R}^i&=&-\xi {\overline \partial} \phi^i-i\xi\chi^j
\Gamma^i_{jm}\lambda_{R}^m~~,
\eea
with $\xi$ a Grassmann odd infinitesimal parameter.
In these equations, $\Gamma$ is  the Levi-Civita connection on $X$.  As in 
\cite{Witten_mirror, Witten_CS}, we define an operator $Q$ through the 
relation:
\be
\delta_Q \Lambda=-i\xi \{Q,\Lambda\}~~,
\ee
where $\Lambda$ is an arbitrary field and $\{Q,\Lambda\}$ stands for the 
graded commutator. 

As  pointed  out  in  \cite{Witten_CS},  one  can  introduce  a  `Floer'
topological  D-brane in  the background.  
This is described  by a  pair $(L,A)$  formed of  a Lagrangian
cycle $L$ and a connection $A$  living in a complex vector bundle $E$
over  $L$.  We  let  $r$   denote  the  (complex)  rank  of  $E$.  
In the presence of the Floer D-brane $(L,A)$, the fields $\phi,\chi,\lambda$ are subject to the boundary conditions:
\bea
\label{bcs}
& &\phi({\partial D})\subset L~~\nn\\
& &\partial_n\phi|_{{\partial D}}~~{\rm~is~a~section~of~}\phi^*(NL)\nn\\
& &\chi|_{{\partial D}} ~~{\rm~is~a~section~of~}\phi^*(TL)~~\\
& &\lambda|_{{\partial D}}~~{\rm~is~form~valued~in~}\phi^*(TL)~~\nn
\eea
Here  we consider  the model  defined on  a disk  $D$, whose boundary we 
denote by $\partial D$.

The  boundary   model  is  obtained  from  its   bulk  counterpart  by
multiplying the following Wilson loop term to the path integrand:
\be
\label{Wilson}
e^{-S_b}:=trPe^{-\int_{{\partial D}}{\phi^*(A)}}
\ee
In  the large  radius limit,  it  was shown  in \cite{Witten_CS}  that
BRST-invariance of the model requires the connection $A$ to be flat. 
As mentioned in the introduction, this conclusion will be 
modified by instanton effects. Therefore, the condition $F_A=0$ should 
be viewed as {\em large radius} string equation of motion, defining 
a semiclassical background in the open string sector
\footnote{By semiclassical we mean that {\em worldsheet} quantum effects are 
being neglected. We shall not treat space-time quantum effects (aka string 
loop effects) in this paper. To avoid confusion, we prefer to use the term 
`large radius background' when referring to this approximation.}.

\subsection{Bulk and boundary observables}
In this subsection, I recall some well-known results on bulk observables,
and re-derive  some properties of the boundary observables
from a `Lagrangian' point of view.

\subsubsection{Bulk observables}

The bulk topological observables of the A-model were constructed in
\cite{Witten_NLSM, Witten_mirror}. The local observables are $Q$-closed
operators                                                        ${\cal
O}_v(z)=v_{I_1...I_k}(\phi(z))\chi^{I_1}(z)...\chi^{I_k}(z)$
associated with closed forms $v$  on the Calabi-Yau manifold $X$. Such
an operator  is large radius BRST-exact
(i.e. exact with respect to the BRST operator appropriate in the large radius 
limit) 
if and  only if the form  $w$ is $d$-exact,
where  $d$ is  the exterior  differential  on $X$.  Hence the  (local)
large radius 
operator BRST  cohomology of the bulk  is isomorphic with  the de Rham
cohomology $H^*(X)$  of the target space $X$.  Beyond these operators,
one  can   build  nonlocal  bulk  observables   by  integrating  their
descendants.  Indeed, it was shown in \cite{DVV} that bulk observables
come in supermultiplets:
\be
\phi(\sigma,\theta)={\cal O}(\sigma)+\theta^\alpha{\cal O}^{(1)}_\alpha(\sigma)+
\theta^{\alpha}\theta^{\beta}{\cal O}^{(2)}_{\alpha\beta}(\sigma)~~,
\ee
with respect to  a bulk $Q$-superspace $(\sigma^\alpha,\theta^\alpha)$
($\alpha=1,2$).  Here  $\sigma^1,\sigma^2$ denote real  coordinates on
the   worldsheet,  related   to   the  complex   coordinate  $z$   via
$z=\sigma^1+i\sigma^2$.   The components  of a  multiplet  are related
through the descent equations:
\bea
d{\cal  O}&=&\{Q,{\cal O}^{(1)}\}~~\\  
d{\cal O}^{(1)}&=&\{Q,{\cal O}^{(2)}\}~~,
\eea
where   we   used   the   worldsheet   forms   ${\cal   O}^{(1)}={\cal
O}^{(1)}_\alpha(\sigma)d\sigma^\alpha$   and   ${\cal   O}^{(2)}={\cal
O}^{(2)}_{\alpha\beta}(\sigma)d\sigma^\alpha\wedge d\sigma^\beta$.

Integrating a  two-form descendant over  the disk produces  a nonlocal
observable whose 
BRST variation is given by\footnote{The  fact that a
boundary  term appears in  the right  hand side  of this  equation was
noticed in \cite{Douglas_quintic}.}:
\be
\{Q,\int_{D}{{\cal O}^{(2)}}\}=\int_{\partial D}{{\cal O}^{(1)}}~~.
\ee
The  nonlocal observables  $\int_{D}{{\cal O}^{(2)}}$  can be  used to
construct topological  bulk deformations \cite{DVV,Hofman},  and their
insertion in  closed sigma model amplitudes produces  topological closed 
string correlators  \cite{DVV,DVV_notes,Dijkgraaf_notes}. This is due
to  the  integral over  the  insertion point  which  is  part of  such
observables' definition.

\subsubsection{Zero-form boundary observables and the boundary BRST operator}

Such operators were constructed in
\cite{Witten_CS}. In fact, only the case of a singly-wrapped D-brane
(when the  bundle $E$ has rank  $r=1$ and $A$ is  a $U(1)$ connection)
was  analyzed  explicitly there,  through the result for  the
nonabelian  case  ($r>1$)  was  mentioned  without  a  detailed
derivation. For completeness, I give the necessary 
derivation in Appendix A. 

A rank $k$ $End(E)$-valued form $w$ on $L$
defines a boundary observable on ${\partial D}$ via:
\be
\label{obs1}
{\cal O}_{w}(x):=w_{\phi(x)}(\chi(x)...\chi(x))~~,
\ee
where $x$ is a point of $\partial D$. Here $w_{\phi(x)}$ is the
alternating multilinear form on $T_{\phi(x)}L$ defined by the value of
$w$ at  the point $\phi(x)$  on $L$. This expression  is geometrically
meaningful since  $\chi(x)$ are sections  of $\phi^*(TL)$, and
is  generally  nonzero even  for  odd  rank  $k$, since  $\chi^I$  are
anti-commuting. Note that  ${\cal O}_w$ can be viewed  as a (Grassmann
even or odd) section of  the bundle $\phi^*(End(E))$. This encodes the
natural   dependence    of   boundary   observables    on   Chan-Paton
factors. 

Let $d_A:\Omega^*(End(E))\rightarrow \Omega^{*+1}(End(E))$ denote the 
covariant differential\footnote{$d_A$ is a differential (i.e. $d_A^2=0$) since $A$ is flat.} with respect to the connection induced by $A$ on 
$End(E)$.
It was argued in \cite{Witten_CS} (and is re-derived by a different method in 
Appendix A) that the following relation holds in the large radius limit:
\be
\{Q_o,{\cal O}_w\}=-{\cal O}_{d_Aw}~~,
\ee\noindent 
where $Q_o$ is the BRST operator in the boundary sector. 
It follows  that  topological boundary  observables  are  
in  one to  one correspondence with elements of $H_{d_A}(L,End(E))$
In particular, the open sector BRST operator can be identified 
with $d_A$ {\em in the large radius limit}. The fact that the 
localized form of $Q_o$ depends on the connection $A$ has far reaching 
implications for the nature of instanton corrections, which will be explored in 
more detail in \cite{sft}. For the moment, it suffices to note that, since 
instanton corrections are known to affect the string field theory action 
\cite{Witten_CS}, the correct string field equations of motion away from the 
large radius limit will require $A$ to obey an equation of the form:
\be
d_AA+{\rm instanton~corrections}=0~~.
\ee
In particular, a flat connection $A$ ($F_A=d_AA=0$) will generally fail to 
satisfy this equation, which is to say that it defines a background 
which does not satisfy the string equations of motion. It
follows that the large radius form $d_A$ of the localized BRST operator 
cannot be correct once instanton effects are taken into account. Since 
the kinetic term of the open string field action is always of the 
form \cite{Thorn}:
\be
S_{kin}=\frac{1}{2}\int{tr(A*Q_o A)}~~,
\ee
this means that the localized string field action of \cite{Witten_CS} 
must contain further corrections induced by worldsheet instanton 
contributions to $Q$. As I show in \cite{sft}(following standard methods 
of string field theory, which were already employed for  
related reasons in \cite{Fukaya, Fukaya2}), these corrections 
can be determined indirectly by first building a non-polynomial string field 
theory around the large radius background given by a flat connection $A$ 
and then shifting this background to a solution of the corrected
string equations of motion. The string field theory action which results after 
this shift consistently takes into account all of the disk instanton 
contributions to the dynamics of the model.

\subsubsection{One-form boundary observables}
Such  operators can  be  built as  descendants  of boundary  zero-form
observables.  BRST-invariance  of the model implies  that the boundary
operators  are  components  of  a $Q_o$-superspace  of  coordinates  $x\in
\partial D$  and $\theta$. This represents  the `boundary restriction'
of  the bulk  superspace $(\sigma^\alpha,\theta^\alpha)$.   It follows
that  the components  of a  boundary  superfield $\Psi(x,\theta)={\cal
O}(x)+\theta {\cal O}^{(1)}(x)$  are related by $d{\cal O}=\{Q_o,{\cal
O}^{(1)}\}$  (the  boundary  descent  equations),  where  $d$  is  the
exterior  differential along  the boundary  and  ${\cal O}^{(1)}={\cal
O}^{(1)}(x)dx$. If ${\cal O}={\cal O}_{w}$ is associated with a closed
form  $w\in \Omega(L,End(E))$,  then it  is  easy to  see that  ${\cal
O}_w^{(1)}(x)=-w_{i_1..i_k}(\phi(x))\phi^{i_1}(x)
\chi^{i_2}(x)...\chi^{i_k}(x)dx$. The {\em trace}
of such operators can  be  integrated  along  a
boundary  segment  $C\subset  \partial   D$  to  produce  a  nonlocal
observable whose BRST-variation is given by:
\be
\{Q_o,\int_{C}{tr({\cal O}^{(1)})}\}=tr({\cal O}(q))-tr({\cal O}(p))~~,
\ee
where $p$ and $q$ are the initial and final points of $C$.

\section{Localization for sigma model open-closed disk amplitudes}

We  are  ready  to  discuss  the geometric  formulation  of  instanton
contributions  to open-closed  amplitudes  on the  disk.  We start  by
considering  sigma model  amplitudes,  and give  largely an  off-shell
treatment,  which   does  not  require   BRST-closure  of  topological
operators.

\subsection{The amplitudes}

We are interested in disk correlators with an arbitrary number of bulk
and  boundary  operator   insertions.  Consider  $n$  forms  $v^{(j)}$
$(j=1..n)$ on $X$ and  $m$ $End(E)$-valued forms $w^{(\alpha)}$
($\alpha=0..m-1)$  on  $L$.  Let  $k_j$ and  $l_\alpha$  denote  their
ranks.  We  shall  give  a  geometric expression  for  the  amplitude:
{\scriptsize \bea
\label{correlator}
\langle \prod_{j}{{\cal O}_{v^{(j)}}(z_j)}
\prod_{\alpha}{{\cal O}_{w^{(\alpha)}}(x_\alpha)}\rangle=~~~~~~~~~~~~
~~~~~~~~~~~~~~~~~~~~~~~~~~~~\nn\\                     
=\frac{1}{r}\int{{\cal
D}[\phi,\chi,\lambda]e^{-S_{bulk}[\phi,\chi,\lambda]}
\prod_{j}{v^{(j)}_{i_1..i_{k_j}}(\phi(z_j))\chi^{i_1}(z_j)...\chi^{i_{k_j}}
(z_j)}\times}~~\\
\times tr\prod_{\alpha}{\left(w^{(\alpha)}
_{i_1...i_{l_\alpha}}(\phi(x_\alpha))
\chi^{i_1}(x_\alpha)...\chi^{l_\alpha}(x_\alpha)U(x_{\alpha},x_{\alpha-1})\right)}~~.\nn
\eea}\noindent Here $I$ and $i$ and indices describing tangent directions to $X$ and $L$, 
and we  order the boundary insertion points  $x_0...x_{m-1}$ cyclically on
$\partial D$ (figure 1).  The  symbol
$U(x_{\alpha},x_{\alpha-1})$   stands   for   the  holonomy   operator
$Pe^{-\int_{x_{\alpha   -1}}^{x_\alpha}{\phi^*(A)}}$  of  $-\phi^*(A)$
along  the  boundary segment  $C_{\alpha-1}=(x_{\alpha-1},x_{\alpha})$
which connects  $x_{\alpha-1}$ to $x_{\alpha}$  in the order  given by
the orientation of $\partial D$.

\iffigs
\hskip 1.0 in
\begin{center} 
\scalebox{0.6}{\begin{picture}(0,0)%
\epsfbox{disk.pstex}%
\end{picture}%
\setlength{\unitlength}{4144sp}%
\begingroup\makeatletter\ifx\SetFigFont\undefined%
\gdef\SetFigFont#1#2#3#4#5{%
  \reset@font\fontsize{#1}{#2pt}%
  \fontfamily{#3}\fontseries{#4}\fontshape{#5}%
  \selectfont}%
\fi\endgroup%
\begin{picture}(3735,3193)(91,-2671)
\put(2116,-1636){\makebox(0,0)[lb]{\smash{\SetFigFont{14}{16.8}{\familydefault}{\mddefault}{\updefault}$z_n$}}}
\put( 91,-1096){\makebox(0,0)[lb]{\smash{\SetFigFont{14}{16.8}{\familydefault}{\mddefault}{\updefault}$x_{m-1}$}}}
\put(406,-2176){\makebox(0,0)[lb]{\smash{\SetFigFont{14}{16.8}{\familydefault}{\mddefault}{\updefault}$C_{m-1}$}}}
\put(2386,-2671){\makebox(0,0)[lb]{\smash{\SetFigFont{14}{16.8}{\familydefault}{\mddefault}{\updefault}$x_0$}}}
\put(3196,-2311){\makebox(0,0)[lb]{\smash{\SetFigFont{14}{16.8}{\familydefault}{\mddefault}{\updefault}$C_0$}}}
\put(3691,-1771){\makebox(0,0)[lb]{\smash{\SetFigFont{14}{16.8}{\familydefault}{\mddefault}{\updefault}$x_1$}}}
\put(3826,-916){\makebox(0,0)[lb]{\smash{\SetFigFont{14}{16.8}{\familydefault}{\mddefault}{\updefault}$C_1$}}}
\put(3646,-106){\makebox(0,0)[lb]{\smash{\SetFigFont{14}{16.8}{\familydefault}{\mddefault}{\updefault}$x_2$}}}
\put(2431,-196){\makebox(0,0)[lb]{\smash{\SetFigFont{14}{16.8}{\familydefault}{\mddefault}{\updefault}$z_2$}}}
\put(1576,-556){\makebox(0,0)[lb]{\smash{\SetFigFont{14}{16.8}{\familydefault}{\mddefault}{\updefault}$z_1$}}}
\end{picture}
}
\end{center}
\begin{center} 
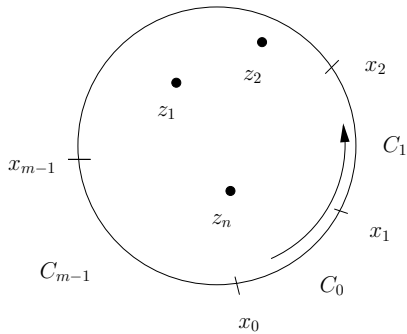
Figure 1.  {\footnotesize Punctured disk associated  to an open-closed
tree-level amplitude.}
\end{center}
\fi

Since some of the observables ${\cal O}$ may
be Grassmann-odd, we  must give a clear ordering  prescription for the
products appearing in  equation (\ref{correlator}).  Our convention is
that  the boundary  product  is taken  in  the opposite  sense of  the
boundary orientation:
\be
\prod_{\alpha}{{\cal O}_{w^{(\alpha)}}(x_\alpha)}:=
{\cal      O}_{w^{(m-1)}}(x_{m-1})...{\cal      O}_{w^{(1)}}(x_1){\cal
O}_{w^{(0)}}(x_0) ~~,
\ee
while the product of bulk insertions is taken in the order from $n$ to
$1$:
\be
\prod_{j}{{\cal O}_{v^{(j)}}(x_j)}=
{\cal        O}_{v^{(n)}}(z_{n})...{\cal        O}_{v^{(2)}}(z_2){\cal
O}_{w^{(1)}}(z_1)~~.
\ee
In (\ref{correlator}), we identify $x_m$  with $x_0$, i.e. we take the
index  of  boundary insertions  to  be  an  element of  $\Z_m$.  Hence
$U(x_0,x_{-1})=U(x_0,x_{m-1})$.  This  amplitude is invariant  {\em up
to  sign} under  an arbitrary  permutation  of bulk  insertions and  
cyclic    permutations   of    boundary    insertions.

\subsection{Localization on instanton configurations}

As  usual   in  topological  field   theory,  the  path   integral  in
(\ref{correlator}) localizes on  the set of $Q$-closed configurations,
which coincide
\cite{Witten_CS,Witten_NLSM,Witten_mirror} 
with with local extrema of the Euclidean action $S$.  This gives a sum
of integrals  over moduli spaces  of instantons.  An instanton  of our
model  is   a  holomorphic  map  $\phi:D\rightarrow   X$  obeying  the
constraint $\phi(\partial  D)\subset L$. The topological  type of maps
from $D$ to $L$ (which plays the role of `instanton class' and acts as
a superselection  index) is given by relative  homotopy classes $\beta
\in  \pi_2(X,L)$
\footnote{For technical reasons this has to be replace with 
$\pi_2^{free}=\pi_0(Map(D,L))$\cite{Kontsevich_recent, Fukaya2}, 
but I will neglect this here. Also, the instanton moduli spaces considered
below may fail to be orientable unless $L$ satisfies extra conditions
\cite{Fukaya2}, another issue which will be ignored below.}.  
For  each  such class,  consider  the moduli  space
${\cal  M}_\beta$ of  all  instantons  which it  contains.  This is  a
non-compact manifold of virtual dimension
\footnote{The virtual dimension is defined as the number of $\chi$ zero modes
minus  the number  of $\lambda$  zero  modes, and  is given  by an  index
theorem discussed in \cite{Fukaya, Fukaya2}.}      $v_\beta=d+\mu(\beta)$,     where
$\mu:\pi_2(X,L)\rightarrow  \Z$  is  the  integer-valued  {\em  Maslow
index} (whose definition is recalled in Appendix B)\footnote{It can be
shown  that the  Maslow index  $\mu(\beta)$  is zero  for all  $\beta$
provided  that the  cycle $L$  is {\em  special} Lagrangian  (the {\em
special} condition means that the  restriction to $L$ of the imaginary
part  of  the  holomorphic  3-form  of $X$  vanishes).   However,  the
open-closed topological  A model makes perfect
sense without  this restriction,  and working with  general Lagrangian
cycles is  important if  one wishes to  study non-BPS D-branes.}.   Note that we  do {\em not}
mod   out  by   the  action   of  $SL(2,\R)$   when   building  ${\cal
M}_\beta$\footnote{In this  respect, we follow the  same convention as
in    \cite{Witten_NLSM,Witten_mirror}   and    \cite{Witten_CS}   for
topological {\em sigma  model} amplitudes. When discussing topological
{\em string}  amplitudes, we  will of course  {\em have to}  divide by
this group.}.

The  value  of  the   Euclidean  action  $S_{bulk}$  on  an  instanton
configuration $\phi$ depends only on its relative homotopy class:
\be
S_{bulk}(\phi)=\int_{D}{\phi^*(\omega)}:=S_{\beta}~~,{\rm~for~all~}\phi~
{\rm~holomorphic~and~of~class~}\beta~~,
\ee
where $\omega$ is the Kahler class of $X$. It follows that:
\be
\langle \prod_{j}{{\cal O}_{v^{(j)}}(z_j)}
\prod_{\alpha}{{\cal O}_{w^{(\alpha)}}}(x_\alpha)\rangle=
\sum_{\beta\in \pi_2(X,L)}{e^{-S_\beta}I_\beta(x,z;v,w)}~~,
\ee
with the quantity $I_\beta$ given by an integral of
\be
\prod_{j}{v^{(j)}_{I_1..I_{k_j}}(\phi(z_j))\chi^{I_1}(z_j)...\chi^{I_{k_j}}
(z_j)}         tr\prod_{\alpha}{\left(w^{(\alpha)}_{i_1...i_{l_\alpha}}
(\phi(x_\alpha))
\chi^{i_1}(x_\alpha)...\chi^{i_{l_\alpha}}(x_\alpha)U(x_{\alpha},x_{\alpha-1})\right)}\nn
\ee
over the  supermanifold associated to ${\cal  M}_\beta$. The fermionic
coordinates  on  this  supermanifold   are  the  $Q$-partners  of  the
infinitesimal variations of a  holomorphic map $\phi$. Since functions
over  this  superspace  can  be  identified  with  forms  over  ${\cal
M}_\beta$, it  follows that $I_{\beta}$  reduces to the integral  of a
certain form over ${\cal M}_\beta$.

\subsection{The universal instanton formalism}

To formulate  localization for  $I_\beta$, it is  convenient to  use a
formalism        similar       to       that        discussed       in
\cite{Witten_NLSM,Witten_mirror}.  Let us  consider the {\em universal
instanton}  $\Phi:D\times {\cal  M}_{\beta}\rightarrow X$,  defined by
$\Phi(z,t):=\phi_t(z)$,  where $\phi_t$  is the  instanton  of modulus
$t\in {\cal  M}_\beta$. In this description, we  view ${\cal M}_\beta$
as a parameter  space for instantons of class  $\beta$.  We will often
identify  $t$ with  $\phi_t$, thereby  writing $\Phi(z,\phi)=\phi(z)$.
We  also   consider  the  restriction  $\Phi_\partial:=\Phi|_{\partial
D\times {\cal M}_\beta}:\partial D\times {\cal M}_\beta
\rightarrow L$. 

The pair  of maps $(\Phi,\Phi_\partial)$ allows us  to pull-back forms
on $X$  and $End(E)$-valued  forms on $L$  to the moduli  space ${\cal
M}_\beta$.   We  analyze this  procedure  for  the  bulk and  boundary
sectors in turn.

\subsubsection{The bulk sector}

Consider a form $v\in  \Omega_k(X)$ and its pull-back $V:=\Phi^*(v)\in
\Omega_k(D\times   {\cal   M}_\beta)$.   The  latter   decomposes   as
$V=V_0+V_1+V_2$, with $V_j\in \Omega_j(D)\otimes
\Omega_{k-j}({\cal M}_\beta)$. 
For each  point $z$ in the  interior of $D$, define  an evaluation map
$e_z:{\cal M}_\beta\rightarrow X$ by:
\be
e_z(\phi):=\phi(z)~~.
\ee
Identifying $\{z\}|\times {\cal M}_\beta$ with ${\cal M}_\beta$, 
this  can be  expressed as  composition  of $\Phi$  with $j_z$,  where
$j_z:\{z\}\times {\cal M}_\beta\rightarrow  D\times {\cal M}_\beta$ is
the inclusion map of the  set $\{z\}\times {\cal M}_\beta$ in $D\times
{\cal     M}_\beta$.    We    are     interested    in     the    form
$V_z:=(e_z)^*(v)=j_z^*(V)=V_0(z)\in \Omega_k({\cal M}_\beta)$.

Now  let us  consider the  case when  $v$ is  closed.   Since exterior
differentiation   on   $D\times    {\cal   M}_\beta$   decomposes   as
$d=d_D+d_{\cal  M}$,   with  $d_D$  and  $d_{\cal   M}$  the  exterior
differentials   on   $D$    and   ${\cal   M}_\beta$,   the   property
$d\Phi^*(V)=\Phi^*(dv)=0$ implies the geometric descent equations:
\bea
\label{descent_bulk}
d_DV_2&=&0~~\nn\\ d_DV_1&=&-d_{\cal M}V_2\\ d_DV_0&=&-d_{\cal M}V_1~~.\nn
\eea
In   this  case,   the  $z$-dependence   of  $V_z$   is   exact,  i.e.
$V_{z'}-V_{z}$ is an exact form on ${\cal M}_\beta$. This follows from
the third  descent equation, upon integration
along a curve $\gamma\subset Int D$ connecting $z$ and $z'$:
\be
V_{z'}-V_z=-d_{\cal M}\int_{\gamma}{V_1}~~.
\ee

\subsubsection{The boundary sector}

Consider  a  rank $l$  form  $w$  on $L$  with  values  in the  bundle
$End(E)$.  The map $\Phi_\partial$ allows us to define the pulled-back
bundle  ${\cal  E}:=(\Phi_\partial)^*(E)$  and  the  pulled-back  form
$W:=(\Phi_\partial)^*(w) \in \Omega_l(\partial D\times {\cal M}_\beta,
End({\cal  E}))$.  Since  $E$ is  flat, ${\cal  E}$ is  a  flat vector
bundle  over $\partial  D\times  {\cal M}_\beta$.  In particular,  the
connection $A$ on  $E$ induces a flat connection  ${\cal A}$ on ${\cal
E}$,  and a covariant  differential $d_{\cal  A}$ on  $End({\cal E})$.
Using the  inclusion map $j_x$  of $\{x\}\times {\cal  M}_\beta$ into
$\partial D\times {\cal M}_\beta$,  we can define `restricted bundles'
${\cal E}(x):=j_x^*({\cal E})$ on ${\cal M}_\beta$, for all points $x$
on the boundary  of the disk, and  evaluation maps $e_x:=\Phi\circ
j_x:{\cal  M}_\beta\rightarrow L$. The  bundles ${\cal  E}(x)$ inherit
flat connections ${\cal A}_x$ induced by restricting ${\cal A}$.

Since the bundle ${\cal E}$ is  flat, the holonomy of $A$ defines maps
${\cal U}((x',t'),(x,t))$  from the  fiber ${\cal E}|_{(x,t)}$  to the
fiber ${\cal E}|_{(x',t')}$ for all  $x,x'$ on the boundary of $D$ and
all $t,t'$  in ${\cal M}_\beta$. These  maps depend only  on the pairs
$(x,t)$  and $(x',t')$.  In  particular, we  have isomorphisms  ${\cal
U}(x',x)(t)={\cal  U}((x',t),(x,t))$  between  ${\cal E}(x)|_{t}$  and
${\cal E}(x')|_{t}$ (see figure 2). 
These give a bundle isomorphism $U(x',x)$ between
${\cal E}(x)$ and ${\cal  E}(x')$. These identifications depend on the
flat connection  $A$ on the topological D-brane.   For our correlator,
we are interested in forms of the type $W_x:=(e_x)^*(w)=(j_x)^*(W)\in
\Omega_l({\cal M}_\beta, End({\cal E}(x)))$.

\

\iffigs
\hskip 1.0 in
\begin{center} 
\scalebox{0.6}{\begin{picture}(0,0)%
\epsfbox{bundles.pstex}%
\end{picture}%
\setlength{\unitlength}{4144sp}%
\begingroup\makeatletter\ifx\SetFigFont\undefined%
\gdef\SetFigFont#1#2#3#4#5{%
  \reset@font\fontsize{#1}{#2pt}%
  \fontfamily{#3}\fontseries{#4}\fontshape{#5}%
  \selectfont}%
\fi\endgroup%
\begin{picture}(4059,2172)(136,-1771)
\put(721,-241){\makebox(0,0)[lb]{\smash{\SetFigFont{12}{14.4}{\familydefault}{\mddefault}{\updefault}$x'$}}}
\put(721,-736){\makebox(0,0)[lb]{\smash{\SetFigFont{12}{14.4}{\familydefault}{\mddefault}{\updefault}$x$}}}
\put(2161,-1771){\makebox(0,0)[lb]{\smash{\SetFigFont{12}{14.4}{\familydefault}{\mddefault}{\updefault}${\cal M}_\beta$}}}
\put(136,-511){\makebox(0,0)[lb]{\smash{\SetFigFont{12}{14.4}{\familydefault}{\mddefault}{\updefault}$\partial D$}}}
\put(2476,-61){\makebox(0,0)[lb]{\smash{\SetFigFont{12}{14.4}{\familydefault}{\mddefault}{\updefault}$(x',t)$}}}
\put(2476,-961){\makebox(0,0)[lb]{\smash{\SetFigFont{12}{14.4}{\familydefault}{\mddefault}{\updefault}$(x,t)$}}}
\end{picture}
}
\end{center}
\begin{center} 
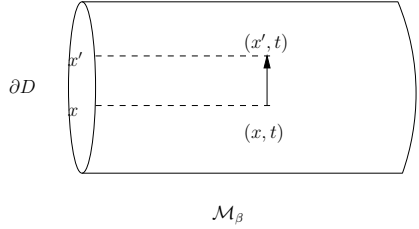
Figure 2. {\footnotesize Two points $(x,t)$ and $(x',t)$ in the direct product
$\partial D\times {\cal M}_\beta$. Parallel transport along the oriented path 
shown in the figure defines the holonomy operator ${\cal U}(x,x')(t)$.}
\end{center}
\fi

Now  consider the  case  when $w$  is  $d_A$-closed on  $L$. Then  the
pulled-back  form $W$  is $d_{\cal  A}$-closed on  $\partial  D \times
{\cal  M}_\beta$,  and  $W_x$  is $d_{{\cal  A}_x}$-closed  on  ${\cal
M}_\beta$.  Moreover, $W$ decomposes as:
\be
W=W_0+W_1~~,
\ee
where $W_0$ and $W_1$ are zero- and one-forms in the 
$\partial D$ directions. In  particular, we  have $W_x=W_0(x)$.  One  can similarly
decompose the connection one-form ${\cal A}$:
\be
{\cal A}={\cal A}_0+{\cal A}_1.
\ee
It is  the easy  to see that  the flatness condition  $d_{\cal A}{\cal
A}=0$ implies that ${\cal A}_0$  and ${\cal A}_1$ are (pull-backs of )
forms on  ${\cal M}_\beta$ and  $\partial D$, respectively  (i.e. they
have  no   dependence  of  the  $\partial  D$   and  ${\cal  M}_\beta$
directions),  and  that  ${\cal   A}_0$  and  ${\cal  A}_1$  are
flat on ${\cal M}_\beta$ and $\partial  D$. The  condition 
$d_{\cal A}W=0$ implies the  geometric boundary
descent equations:
\bea
d_{{\cal A}_1}W_1&=&0~~\nn\\
d_{{\cal  A}_1}W_0&=&-d_{{\cal A}_0}W_1~~\\
d_{{\cal  A}_0}W_0&=&0~~.\nn\\ 
\eea

\subsection{Localization formula for $I_\beta$}

With  these  preliminaries, we  are  ready  to  give the  localization
formula for $I_\beta$: {\scriptsize\be
\label{loc_nlsm}
I_\beta=
\frac{1}{r}\int_{{\cal M}_\beta}{V^{(n)}_{z_n}\wedge ... \wedge V^{(1)}_{z_1}
tr\left[W^{(m-1)}_{x_{m-1}}{\cal
U}(x_{m-1},x_{m-2})W^{(m-2)}_{x_{m-2}}                            {\cal
U}(x_{m-2},x_{m-3})W^{(m-3)}_{x_{m-3}}....{\cal  U}(x_1,x_0)W^{(0)}_{x_0}  {\cal
U}(x_0,x_{m-1})\right]}~~,~~~~~~~~~~~~~~~~~~~~~~~~~~~~~~
\ee}\noindent where $r$ denotes the rank of $E$. 
This relation can be justified though path integral 
arguments similar to those of \cite{Witten_NLSM, Witten_CS}.

\subsection{Ghost grading and the ghost number anomaly}

The  open-closed nonlinear  sigma  model has  a  global $U(1)$  `ghost
number' symmetry under which  $\phi,\chi$ and $\lambda$ have charges $0,+1$ and
$-1$  respectively.   In  particular,  the  bulk  and  boundary  local
observables ${\cal  O}_v$, ${\cal O}_w$  have charges $rank  v$ and $rank
w$.  In the  quantum theory,  this symmetry  is anomalous  due  to the
nontrivial geometric character of the Grassmann-odd worldsheet fields. 
In  a bosonic
background configuration $\phi$, the  anomaly is computed by the index
of  the  associated  Dirac  operator,  which is  proportional  to  the
difference between the  number of $\chi$ zero modes  and the number of
$\lambda$  zero   modes.  This  follows  by  the   standard  argument  of
\cite{Witten_NLSM}.   The   result  is  given  by   an  index  theorem
\cite{Fukaya, Fukaya2},  on  surfaces  with
boundaries which  for the  case of disks  gives the  virtual dimension
$d+\mu(\beta)$  mentioned above.  The contribution $d=\frac{1}{2}dim_{\R}X$
is the standard  ghost number anomaly on the  disk, while $\mu(\beta)$
appears as a  correction due to the nontrivial  boundary conditions on
the fermions.

Knowledge  of the ghost  number anomaly  allows us  to write  down the
selection  rules   for  amplitudes.  Namely,  the instanton
contribution $I_\beta$ vanishes unless:
\be
\sum_{j}{rank v_j}+\sum_{\alpha}{rank w_\alpha}=d+\mu(\beta)~~.
\ee 
This  follows  immediately  from  ghost number  conservation  or  more
directly from the localization formula (\ref{loc_nlsm}).

\subsection{Exceptional sigma model amplitudes}

Let  us  take a  closer  look at  the  disk  amplitudes containing  one or 
two boundary  insertions or a  single  bulk
insertion. These  have certain non-renormalization properties,  
which originate from the fact that  specifying the position of such  
sets of punctures does
not  suffice to  completely  fix the  gauge  under the the
$SL(2,\R)$ symmetry of the disk.

\subsubsection{The two-point boundary amplitude}

This    is    the    sigma    model    correlator    $\langle    {\cal
O}_{w^{(1)}}(x_1){\cal O}_{w^{(0)}}(x_0)\rangle=
\sum_{\beta}{I_\beta e^{-S_\beta}}$, with: 
\be
\label{I}
I_\beta=\frac{1}{r}\int_{{\cal M}_\beta}{tr\left[W^{(1)}_{x_1}{\cal
U}(x_1,x_0)W^{(0)}_{x_0}{\cal U}(x_0,x_1)
\right]}~~.
\ee
It is  easy to see that the  obvious action of $SL(2,\R)$ 
has a  one-dimensional subgroup $G$  which fixes the points  $x_0$ and
$x_1$. This is  isomorphic with the  additive group  $\R$ of
real numbers. If  one maps the disk to the upper  half plane such that
$x_0$ and  $x_1$ are mapped to the  origin $0$ and the  unit $1$, then
the elements of $G$ are $SL(2,\R)$ matrices of the form:
\be
A_\xi=\left[\begin{array}{cc}e^\xi&0\\2sinh(\xi)&e^{-\xi}\end{array}\right]~~,
\ee
which obey $A_{\xi_1}A_{\xi_2}=A_{\xi_1+\xi_2}$ and $A_0=Id$.  In this
presentation,                        an                        element
$A=\left[\begin{array}{cc}a&b\\c&d\end{array}\right]
\in SL(2,\R)$ acts on the upper half plane via $z\rightarrow f_A(z)=\frac{az+b}{cz+d}$. 
For ease of notation, we denote $f_{A_\xi}$ simply by $f_\xi$.

The transformations  $f_\xi$ act naturally on the  moduli space ${\cal
M}_\beta$. If  $\phi$ is  an instanton in  ${\cal M}_\beta$,  then the
action of  $A_\xi$ takes it  into $\phi\circ f_{-\xi}$.  The composite
map is still an instanton, since $f_\xi$ is holomorphic. We denote the
resulting action of $\R$ on ${\cal M}_\beta$ by $\rho$, i.e.  we write
$\rho_\xi(\phi):=\phi\circ  f_{-\xi}$.  Since  $f_\xi$ fixes  the points
$x_0$ and  $x_1$, such transformations  of ${\cal M}_\beta$  leave the
evaluation maps unchanged:
\be
e_{x_\alpha}\circ \rho_{\xi}=e_{x_\alpha}~~{\rm~for~}\alpha=0,1~~.
\ee
Indeed,  one  has $e_{x_\alpha}(\rho_\xi(\phi))=e_{x_\alpha}(\phi\circ
f_{-\xi})=(\phi\circ f_{-\xi})(x_\alpha)=
\phi(f_{-\xi}(x_\alpha))=\phi(x_\alpha)=e_{x_\alpha}(\phi)$. This invariance property implies that 
$e_{x_\alpha}$  descend  to   evaluation  maps  ${\hat  e}_{x_\alpha}$
defined  on  the quotient  ${\hat  {\cal M}}_\beta={\cal  M}_\beta/\R$. 
If $\pi:{\cal M}_\beta
\rightarrow {\hat {\cal M}}_\beta$ is the associated projection, 
the induced maps satisfy 
$e_{x_\alpha}={\hat  e}_{x_\alpha}\circ  \pi$,  which shows  that  the
forms   $W^{(\alpha)}   _{x_\alpha}=e^*_{x_\alpha}(w^{(\alpha)})$  are
basic on this fibration:
\be
W^{(\alpha)}_{x_\alpha}=\pi^*(\omega^{(\alpha)}_{x_\alpha})~~,
\ee
with  $\omega^{(\alpha)}_{x_\alpha}={\hat  e}^*_{x_\alpha}(w^{(\alpha)})$ some 
forms on ${\hat {\cal M}}_\beta$ taking values in the bundles $End({\hat {\cal
E}}_{x_\alpha})$,    where    ${\hat    {\cal    E}}_{x_\alpha}:={\hat
e}^*_{x_\alpha}(E)$. 

We  can  now prove  a  non-renormalization  theorem  for the  
two-point boundary amplitude. 
For this, consider the case when the homotopy class 
$\beta$ is nontrivial. 
In this situation,  the action $\rho$ on ${\cal  M}_\beta$ reduces its
dimension, i.e. the quotient ${\hat {\cal M}}_\beta$ has dimension one
unit  less than  ${\cal M}_\beta$.  It immediately  follows 
that  the integral (\ref{I}) must vanish
\footnote{When performing this integral one has to evaluate the integrand on 
bases $u_1..u_D$ of tangent vectors to ${\cal M}_\beta$, where 
$D=dim{\cal M}_\beta$; this has to be done on a coordinate cover of 
${\cal M}_\beta$ subordinate to a partition of unity, according to the definition of 
integration of forms. One can always assume that $u_1$ is in the direction 
of the $\R$ fiber, which implies that $\pi_*(u_1)=0$. Since both forms involved in 
(\ref{I}) are basic, their values on  $u_j$ can be expressed in terms of 
$\pi_*(u_j)$, hence the value of the integrand on $u_1..u_D$ is always zero.}.
The  case $\beta=0$  (the  trivial
homotopy class) is different. In  this situation, the space ${\cal M}_0$
coincides  with $L$,  since area-minimizing  maps of  trivial relative
homotopy must be constant on the  disk $D$. Since the action $\rho$ is
trivial on  such  maps, the  quotient  ${\hat {\cal  M}}_0$
coincides  with  ${\cal  M}_0$  and the argument given
above can  no longer be applied. It  follows that the entire
contribution  to  the  two-point  amplitude  comes  from  the  trivial
instanton sector, and can be expressed geometrically as an integral over
$L$:
\be
\langle {\cal O}_{w^{(1)}}{\cal O}_{w^{(0)}}\rangle=I_0=
\frac{1}{r}\int_{L}{tr\left[w^{(1)}\wedge w^{(0)}\right]}~~.
\ee
Since this quantity localizes exclusively on the trivial instanton sector, 
it does not depend on the insertion points $x_0$ and $x_1$. 

The  two-point {\em  sigma model}  correlator on  the disk  defines the
so-called  {\em boundary topological  metric}:
\be
\label{rho}
\rho(w_1,w_2):=\langle {\cal O}_{w_1}{\cal O}_{w_2}\rangle~~.
\ee 
Hence  we recover  the result
that {\em  the boundary topological metric does  not receive instanton
corrections}. A  different argument  to the same  effect 
can be extracted from \cite{Hofman}. However, the approach of \cite{Hofman} 
assumes a background which is conformally invariant and satisfies the equations 
of motion, which is precisely the assumption we wish to avoid.

\subsubsection{The one-point bulk amplitude on the disk}

A  similar   non-renormalization  result  can  be   derived  for  disk
amplitudes $\langle{\cal O}_v(z)\rangle$ with a single bulk insertion.
In this case, the stabilizer  of the insertion point $z$ in $SL(2,\R)$
is isomorphic with the rotation group $SO(2)$. Mapping the disk to the
upper half-plane such that $z$  is mapped into the imaginary unit, the
$SL(2,\R)$ matrices which stabilize $i$ have the form:
\be
A_\theta=\left[\begin{array}{cc}\cos\theta                        &\sin
\theta\\-\sin\theta&\cos\theta
\end{array}\right]~~,
\ee
with $\theta\in \R/(2\pi\Z)$. As before, we denote their action on the
disk by $f_\theta=f_{A_\theta}$. One  has the expansion $\langle {\cal
O}_v(z)\rangle=\sum_{\beta}{I_\beta e^{-S_\beta}}$, with:
\be
\label{Ibulk}
I_\beta=\frac{1}{r}\int_{{\cal M}_\beta}{H_AV_{z}}~~,
\ee
where $H_A$ is a  function on
${\cal M}_\beta$ given by the fiber-wise trace of the holonomy operator
${\cal U}(x-,x+):{\cal E}(x)
\rightarrow {\cal E}(x)$ associated to the path which connects a point 
$x\in \partial D$  with itself after winding once  around the boundary
of the disk  (figure 3). The fiber-wise trace $H_A$
does not  depend on the  choice of $x$,  since changing $x$  to $x'\in
\partial D$ amounts to a similarity transformation
\footnote {One can in fact identify $H_A(\phi)$ with the spacetime holonomy 
$W(\phi)=\frac{1}{r}trPe^{-\int_{\partial D}{\phi^*(A)}}$, which depends only on the 
homotopy class $\beta$; this allows one to write $I_\beta=W_\beta\int_{{\cal M}_\beta}{V_z}$.
Since $I_\beta$ will localize on the trivial instanton sector, this  
turns out to be irrelevant.}:
\be
{\cal U}(x'-,x'+)={\cal U}(x',x){\cal U}(x-,x+){\cal U}(x',x)^{-1}~~.
\ee

\iffigs
\hskip 1.0 in
\begin{center} 
\scalebox{0.6}{\begin{picture}(0,0)%
\epsfbox{hol.pstex}%
\end{picture}%
\setlength{\unitlength}{4144sp}%
\begingroup\makeatletter\ifx\SetFigFont\undefined%
\gdef\SetFigFont#1#2#3#4#5{%
  \reset@font\fontsize{#1}{#2pt}%
  \fontfamily{#3}\fontseries{#4}\fontshape{#5}%
  \selectfont}%
\fi\endgroup%
\begin{picture}(6528,1962)(910,-1672)
\put(3041,-915){\makebox(0,0)[lb]{\smash{\SetFigFont{12}{14.4}{\familydefault}{\mddefault}{\updefault}$x+$}}}
\put(3009,-1202){\makebox(0,0)[lb]{\smash{\SetFigFont{12}{14.4}{\familydefault}{\mddefault}{\updefault}$x$}}}
\put(2849,-1426){\makebox(0,0)[lb]{\smash{\SetFigFont{12}{14.4}{\familydefault}{\mddefault}{\updefault}$x-$}}}
\put(4501,-1231){\makebox(0,0)[lb]{\smash{\SetFigFont{12}{14.4}{\familydefault}{\mddefault}{\updefault}$x$}}}
\put(6481,-1051){\makebox(0,0)[lb]{\smash{\SetFigFont{12}{14.4}{\familydefault}{\mddefault}{\updefault}$t$}}}
\end{picture}
}
\end{center}
\begin{center} 
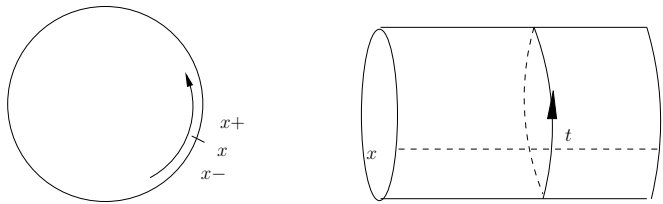
Figure 3. {\footnotesize Parallel around the disk boundary induces 
an automorphism of the bundle ${\cal E}(x)$ defined over ${\cal M}_\beta$.}
\end{center}
\fi

Now consider  the   action  $\rho_\theta(\phi):=\phi\circ
f_{-\theta}$ of $SO(2)$ on ${\cal  M}_\beta$ and notice as before that
it preserves the evaluation map at $z$:
\be
e_z\circ \rho_\theta=e_z~~.
\ee 
Hence $e_z$  descends to  a map ${\hat  e}_z$ defined on  the quotient
${\hat   {\cal   M}}_\beta={\cal   M}_\beta/SO(2)$,  which   satisfies
$e_z={\hat e}_z\circ  \pi$ with  respect to the  associated projection
$\pi$. This implies that the form $V_z$ is basic:
\be
V_z=\pi^*(\omega)~~,
\ee
with  $\omega$ a  form on  ${\hat {\cal  M}}_\beta$. 
As in the previous subsection, we conclude that
$I_\beta=0$ unless $\beta$ is  the trivial homotopy class. Hence
the bulk  one-point functions  on the disk  
are given exactly by their large radius (trivial
instanton sector) expression:
\be
\label{bb}
\langle {\cal O}_v\rangle= \int_{L}{v}~~.
\ee
To obtain the  last formula, we used that  fact that ${\cal M}_0=L$
and noticed that the holonomy  of $E$ around a curve on $L$
which  is  homotopically trivial  in  $L$  is  given by  the  identity
operator
\footnote{What we are considering here is a coupling 
of the  A-model boundary state  to bulk {\em A-model}  operators.  One
should not confuse such correlators with the couplings of the boundary
A-model vacuum  to {\em B-model}  observables. The latter have  a very
different character  and are, in particular, responsible  for the mass
of wrapped D-branes
\cite{Yakov,branes1,branes2,Moore_dimension}. In particular, our 
non-renormalization arguments say nothing about such couplings.}.

\subsubsection{The one-point boundary amplitude on the disk}

Finally, we consider one-point boundary amplitudes 
$\langle {\cal O}_w(x)\rangle$. A similar argument shows that such correlators
do not receive instanton corrections, and hence are given by their 
large radius value:
\be
\langle {\cal O}_w\rangle =\frac{1}{r}\int_{L}{tr(w)}~~.
\ee
This allows us to express them in terms of the 
boundary topological metric:
\be
\langle {\cal O}_w\rangle =\rho(w,1_E)=\langle{\cal O}_w{\cal O}_{1_E}\rangle~~, 
\ee
where $1_E$ is the identity endomorphism of $E$.

\section{Open-closed string amplitudes on the disk}

\subsection{Preliminary remarks}

A  complete  study of  the  open-closed  string  theory based  on  the
topological  A-model requires  that we  consider its  coupling  to the
open-closed    version   of   two-dimensional    topological   gravity
\cite{Witten_TG}.   Though  the   present  paper   does   not  discuss
open-closed topological gravity,  it is is not hard  to understand the
final  effect of  the  construction.  As in  the  closed string  case,
coupling  to  topological  gravity  plays  the  role  of  implementing
integration  over worldsheet  metrics.  In fact,  the resulting  model
contains  the  two-dimensional metric  as  a  dynamical variable,  and
diffeomorphism invariance is recovered only after integrating over all
such  metrics. As  in the  closed string  case, this  can  be formally
encoded  in   the  statement  that  bulk  and   boundary  sigma  model
observables  are promoted  to string  observables  upon multiplication
with puncture operators. The  later are local operators of topological
gravity, associated  to punctures on  a given Riemann surface.  In the
open-closed  case,   these  operators   come  in  bulk   and  boundary
incarnations, which  we denote  by $P(z)$ and  $P(x)$. Given  bulk and
boundary sigma model observables  ${\cal O}_v(z)$ and ${\cal O}_w(x)$,
the associated string operators are:
\bea
{\hat    {\cal    O}}_v(z)&=&{\cal    O}_v(z)P(z)~~\\   {\hat    {\cal
O}}_w(x)&=&{\cal O}_w(x)P(x)~~,
\eea
and we are interested in amplitudes containing such insertions:
\be
\langle\langle \prod_{j}{{\cal O}_{v^{(j)}}}\prod_{\alpha}{
{\cal O}_{w^{(\alpha)}}}\rangle\rangle:=
\langle \prod_{j}{{\hat {\cal O}}_{v^{(j)}}}\prod_{\alpha}{
{\hat {\cal O}}_{w^{(\alpha)}}}\rangle_{string}~~.
\ee
Inclusion  of  $P(z)$  and   $P(x)$  is  necessary  for  a  consistent
implementation of the integral over puncture positions.

In  fact, this description  is somewhat  simplified, since  it assumes
that  we  have chosen  the  `puncture  representation' of  open-closed
Riemann  surfaces.   Following  standard  procedures of  string  field
theory
\cite{Zwiebach_closed,Zwiebach_open}, one can work with a 
more general  description, which uses a  different parameterization of
the  associated  moduli space.  It  is  well-known  from the  operator
(Segal)   formalism   of   conformal   field   theories   \cite{Segal,
operator_formalism} that the geometric objects necessary for a correct
description  of  conformal  field  theory  are  punctured  open-closed
Riemann  surfaces together  with a  choice of  holomorphic coordinates
around   the   punctures.   Such    objects   are   elements   of   an
infinite-dimensional  moduli  space, which  projects  onto the  moduli
space  of uncoordinatized  punctured  surfaces through  the map  which
forgets the coordinates around  all punctures. The correct description
of Polyakov  amplitudes is  through integrals of  certain differential
forms    over    finite-dimensional    chains    defined    on    this
infinite-dimensional moduli space.  The precise choice of these chains
amounts to a  specification of string vertices, and  is constrained by
certain  consistency  conditions (the  geometric  vertex equations  of
\cite{Zwiebach_closed,  Zwiebach_open}) which express  the requirement
that the domain  of integration in string perturbation  theory gives a
single-cover  of   the  moduli  space   of  uncoordinatized  punctured
surfaces.  The  `puncture  representation'  arises from  a  particular
choice of  open-closed string vertices, which consists  of taking them
to   carry  infinite-length   stubs  and   strips.  This   is  related
\cite{Distler_Nelson}to the conformal normal ordering prescription of
\cite{Polchinski_hol_gauge} used in the traditional description of 
string perturbation theory.

The purpose of recalling these  well-known results is to stress
that all string  amplitudes considered in this paper  are taken in the
puncture representation, hence our string vertices carry infinite length
stubs  and  strips  associated   with  each  open  and  closed  string
insertion.  This is by no  means the most general consistent approach,
and other  choices are more  suitable for foundational studies  or for
the  rigorous  definition  of  the  associated  string  measure.  

Ability to write Polyakov amplitudes  as integrals over a moduli space
of  complex  Riemann  surfaces   (i.e.  as  integrals  over  conformal
equivalence  classes   of two-dimensional  metrics)   depends  on  the
assumption  that  the  underlying   sigma  model  is  {\em  off-shell}
conformally invariant, i.e. is  a `conformal topological field theory'
in  the   sense  of  \cite{DVV}.   Many of the   standard  geometric
manipulations  of  string  perturbation  theory are justified  only in the
presence of  off-shell conformality. In particular,  the equivalence of
amplitudes defined  by two conformally equivalent  Riemann surfaces is
not correct unless off-shell conformality holds--for example, there is
no way to  identify the amplitudes on a surface  with finite size open
or  closed  string  boundaries  with  the  amplitude  defined  on  the
conformally  equivalent punctured surface  (which  can be visualized as 
carrying infinitely long  stubs and strips at the insertions).  
Much of the geometric  discussion of localized
string  amplitudes presented  below  can be  carried  out without  the
off-shell conformality assumption, but we wish to warn the reader that
this is  a mathematical artifact due  to the fact that  we tacitly use
the  puncture  representation  when  writing all  string  localization
formulae\footnote{No such restriction is  needed for {\em sigma model}
amplitudes, which make perfect physical sense without the conformality
assumption.  In  fact, one could consider the  open-closed sigma model
on almost  Kahler manifolds, by  performing the boundary  extension of
the  closed  model of  \cite{Witten_NLSM}.  Though  in  this paper  we
consider the Calabi-Yau model  for simplicity (which assures off-shell
conformal invariance of  the sigma model in the  {\em closed} sector),
there  is  no need  to  impose  conformal  invariance when  discussing
open-closed sigma  model amplitudes, and  we have avoided doing  so in
previous sections.   Conformal invariance  in the open  sector 
can be assured by restricting to special Lagrangian cycles $L$.} .  
One  must assume off-shell conformal invariance if
one wishes  to interpret these  objects as standard  string amplitudes
obeying  the  axiomatic  requirements  of  \cite{Zwiebach_open}.

Let me add  a few observations on open-closed  topological gravity. As
in the  closed case, pure  topological gravity on  open-closed Riemann
surfaces contains many more  observables beyond the puncture operators
$P(z)$  and $P(x)$.  These  can be  used to  define bulk  and boundary
{\em gravitational}  descendants  of our  `primary'  operators ${\hat  {\cal
O}}_v(z)$  and ${\hat  {\cal O}}_w(x)$.  In particular, one 
could use field theory arguments in order to propose recursion 
relations between string amplitudes of primaries (to be discussed below) 
and amplitudes containing gravitational descendants (which will not be 
discussed in this paper). 
In  the closed  case, 
pure  topological  gravity and  topological  gravity
coupled  to $c<1$  matter  are integrable  models,  related to  matrix
models and KdV hierarchies  \cite{TG_integrable}. It is an interesting
question  to what extend  such results  generalize to  the open-closed
case. To my knowledge, little work has been done in this direction.

\subsection{Moduli spaces of punctured disks}

The localization formula for string amplitudes involves integrals over
a  moduli space  of  instantons associated  with  punctured disks.  To
define these, consider the unit disk $D$ together with $n$ bulk points
$z_j$  and  $m$ boundary  points  $x_\alpha$,  as  in Subsection  2.1.
Mapping the  disk to the upper  half-plane (such that  $\partial D$ is
mapped  into  the  real  axis),  we  have  $SL(2,\R)$  transformations
$z\rightarrow  f_A(z)=\frac{az+b}{cz+d}$   induced  by  real  matrices
$A=\left(\begin{array}{cc}a&b\\c&d\end{array}\right)$      of     unit
determinant.  Such  transformations preserve  the  the orientation  of
$\partial  D$, since the  function $f_A$  is strictly  increasing when
restricted  to the  real axis.  In particular,  the  $SL(2,\R)$ action
preserves  the cyclic  ordering of  the points  $x_\alpha$  around the
boundary.

We define a moduli space ${\cal M}(\beta)={\cal M}^n_m(\beta)$ of maps
$\phi$ associated with  bulk-boundary punctured disks. More precisely,
since the  un-punctured disk  $D$ has no  moduli, the moduli  space of
punctured  disks with $n$  bulk and  $m$ boundary  punctures coincides
with   the  moduli   space   of  bulk   and  boundary   configurations
$z=(z_1..z_n)\in (Int D)^n$ and $x=(x_0..x_{m-1})\in (\partial D)^n_{cycl}$,
where  we consider  only  boundary configurations  which preserve  the
cyclic order  on $\partial D$. This  moduli space can  be described as
the quotient of the set of triples $(z,x,\phi)$ via the obvious action
of $SL(2,\R)$:
\bea
\label{SL_action}
z&=&(z_1..z_n)\rightarrow                      (f_A(z_1)..f_A(z_n))~~\nn\\
x&=&(x_0..x_{m-1})\rightarrow (f_A(x_0)..f_A(x_{m-1}))~~\\
\phi&\rightarrow& \phi\circ f_A^{-1}~~.\nn
\eea
Since  we   `integrate'  over   puncture  positions  (and   divide  by
$SL(2,\R)$),   the  real   dimension  of   ${\cal   M}(\beta)$  equals
$d+2n+m-3+\mu(\beta)$ (for the moment we  exclude cases  with $n+m<3$). 
It  will prove
useful to  also consider the  moduli space ${\tilde  {\cal M}}(\beta)=
{\tilde  {\cal M}}_m^n(\beta)$  of triples  $(z,x,\phi)$  {\em before}
performing the $SL(2,\R)$ quotient. This is a direct product:
\be
{\tilde {\cal M}}(\beta)=S_n\times C_m\times {\cal M}_\beta~~,
\ee
where         $S_n=\{(z_1...z_n)\in         (Int         D)^n|z_1..z_m
{\rm~are~all~distinct}\}$   and   $C_m:=\{(x_0..x_{m-1})\in  (\partial
D)^m|x_0...x_{m-1}        {\rm       ~are~all~distinct~and~cyclically~
ordered~along~}\partial  D\}$. The moduli  space ${\cal  M}(\beta)$ is
the  quotient of  ${\tilde  {\cal M}}(\beta)$  through the  $SL(2,\R)$
action (\ref{SL_action}).

Finally,  we  consider  bulk  and boundary  evaluation  maps  ${\tilde
e}_j:{\tilde    {\cal    M}}(\beta)\rightarrow    X$    and    ${\tilde
e}_\alpha:{\tilde {\cal M}}(\beta)\rightarrow L$ defined through:
\bea
{\tilde             e}_j(z,x,\phi)&:=&\phi(z_j)~~\\            {\tilde
e}_\alpha(z,x,\phi)&:=&\phi(x_\alpha)~~.
\eea
These     are     obviously     invariant     under     the     action
(\ref{SL_action}).  Hence  they  induce well-defined  maps  $e_j:{\cal
M}(\beta)\rightarrow X$  and $e_\alpha:{\cal M}(\beta)\rightarrow  L$ on
the  quotient space.   If we  let  $p:{\tilde {\cal
M}}(\beta)\rightarrow {\cal M}(\beta)$  denote the obvious projection,
then we have:
\be
\label{ep}
{\tilde e}_j=e_j\circ p~~,~~{\tilde e}_\alpha=e_\alpha\circ p~~.
\ee

The maps  $e_\alpha,{\tilde e}_\alpha$ allow us  to define pulled-back
bundles    ${\cal   E}_\alpha:=e_\alpha^*(E)$    ,    ${\tilde   {\cal
E}}_\alpha={\tilde  e}_\alpha^*(E)$   on  the  moduli   spaces  ${\cal
M}(\beta)$  and ${\tilde  {\cal M}}(\beta)$.   If $u=(z,x,\phi)$  is a
point  in ${\tilde  {\cal  M}}(\beta)$, the  fiber  of ${\tilde  {\cal
E}}_\alpha$   at   $u$   coincides   with   the  fiber   of   $E$   at
$\phi(x_\alpha)$.  The   flat  connection  on   $E$  defines  parallel
transport  operators  $U(\phi(x_{\alpha_2}),\phi(x_{\alpha_1}))$  from
$E_{\phi(x_{\alpha_1})}$  to  $E_{\phi(x_{\alpha_2})}$,  which  induce
bundle   isomorphisms  ${\tilde  {\cal   U}}_{\alpha_2\alpha_1}$  from
${\tilde      {\cal     E}}_{\alpha_1}$     to      ${\tilde     {\cal
E}}_{\alpha_2}$. Equations (\ref{ep})  imply that the bundles ${\tilde
{\cal  E}}_\alpha$ are pull-backs  through the  projection $p$  of the
bundles ${\cal  E}_\alpha$. In fact, it  is easy to see  that the maps
${\tilde  {\cal  U}}_{\alpha_2\alpha_1}$  are also  well-behaved  with
respect to this projection, i.e. they descend to isomorphisms 
${\cal U}_{\alpha_2\alpha_1}$   from    ${\cal   E}_{\alpha_1}$   to   ${\cal
E}_{\alpha_2}$.  The latter  give us  a  way to  identify the  various
bundles ${\cal  E}_\alpha$, which will be crucial  when writing string
amplitudes.  The identifications ${\cal  U}_{\alpha_2\alpha_1}$ depend
on the flat connection $A$ on the cycle $L$.

\subsection{Localization formula for string amplitudes}

We  consider  the  (disk)  string amplitude  ${\cal  B}=\langle\langle
\prod_{j}{{\cal O}_{v^{(j)}}(z_j)}
\prod_{\alpha}{{\cal O}_{w^{(\alpha)}}(x_\alpha)}\rangle \rangle$, where 
$v^{(j)}$, $w^{(\alpha)}$  are forms on  $X$ and $L$ as  in Subsection
2.1.1. Since we are  interested in working off-shell, we  do {\em not}
require these  forms to  be closed. The  topological character  of the
model implies that such  amplitudes localize on disk instantons. Hence
one has an expansion:
\be
{\cal B}(z,x;v,w)=\sum_{\beta\in \pi_2(X,L)}{e^{-S_{\beta}}J_\beta(z,x,v,w)}~~.
\ee
\noindent We claim that the localization formula for $J_\beta$ is:
\be
\label{loc_string}
J_\beta=\frac{1}{r}\int_{{\cal
M}(\beta)}{\left[\wedge_{j=n}^1{e_j^*(v^{(j)})}\right]
\wedge tr\left[
\wedge_{\alpha=m-1}^{0}{\left(e_\alpha^*(w^{(\alpha)}){\cal U}_{\alpha,{\alpha-1}}\right)}\right]}~~,
\ee
where ${\cal  M}(\beta)$ and $e_j,e_\alpha$ are the  moduli spaces and
bulk/boundary evaluation maps defined  in the previous subsection.  In
this  equation,  $e_j^*(v^{(j)})$  and $e_\alpha^*(w^{(\alpha)})$  are
complex-valued,  respectively $End({\cal  E}_\alpha)$-valued  forms on
${\cal M}_\beta$,  while $tr$  denotes the fiber-wise  trace.  Equation
(\ref{loc_string})  can be  justified through  a standard  analysis of
path integral  localization.

Let us write the ghost number selection rule for $J_\beta$:
\be
\sum_{j=1}^n{rank v_j}+\sum_{\alpha=0}^{m-1}{rank w_\alpha}=
dim{\cal M}^n_m(\beta)=d+2n+m-3+\mu(\beta)~~.
\ee
In the conformal case we have $\mu(\beta)=0$ for all $\beta$, which gives a
global selection rule for the disk string amplitude ${\cal B}$:
\be
\sum_{j=1}^n{rank v_j}+\sum_{\alpha=0}^{m-1}{rank w_\alpha}=d+2n+m-3~~.
\ee

\subsection{Relation of string amplitudes to sigma model amplitudes of 
descendants}

We  would like to  give an  open-closed version  of the  closed string
argument
\cite{Dijkgraaf_notes} that nonlinear sigma model amplitudes with an appropriate 
number  of integrated  descendants reproduce  string  amplitudes. More
precisely, we  wish to show that the  instanton contribution $J_\beta$
to the string  amplitude ${\cal B}$ has one  the following alternate
expressions:

{\tiny
\bea
\label{string1}
J_\beta&=&~~
\int_{S_{n-1}(z_1)\times C_{m-1}(x_0)}{
\int^{(\beta)}{{\cal D[\phi,\chi,\lambda]}{\left[\wedge_{j=n}^2{{\cal O}^{(2)}_{v^{(j)}}}\right] {\cal O}_{v^{(1)}}(z_1)}}}~~\nn\\
~&\wedge &tr\left(
\wedge_{\alpha=m-1}^1{\left[{\cal O}^{(1)}_{w^{(\alpha)}}(x_{\alpha})U(x_\alpha,x_{\alpha-1})\right]} 
{\cal O}_{w^{(0)}}(x_0)U(x_0,x_{m-1})\right)=\\
&=&\int_{S_{n-1}(z_1)\times C_{m-1}(x_0)}
\int_{{\cal M}_\beta}{\left[\wedge_{j=n}^2{V^{(j)}_2}\right]
\wedge V^{(1)}_0(z_1)}\wedge tr\left(
\wedge_{\alpha=m-1}^1{\left[W^{(\alpha)}_1{\cal U}(x_\alpha,x_{\alpha-1})\right]} \wedge
W^{(0)}_0(x_0){\cal U}(x_0,x_{m-1})\right)~~,\nn
\eea}
{\scriptsize\bea
\label{string2}
J_\beta&=&
\int_{S_{n}\times C_{m-3}(x_0,x_1,x_2)}
{\int^{(\beta)}{
{\cal                      D[\phi,\chi,\lambda]}\left[\wedge_{j=n}^1{{\cal
O}^{(2)}_{v^{(j)}}}\right]}}\nn\\
~&\wedge &tr\left(
\wedge_{\alpha=m-1}^3{\left[{\cal O}^{(1)}_{w^{(\alpha)}}(x_{\alpha})
U(x_\alpha,x_{\alpha-1})\right]} 
{\cal O}_{w^{(2)}}(x_2)U(x_2,x_1) {\cal O}_{w^{(1)}}(x_1)U(x_1,x_0){\cal
O}_{w^{(0)}}(x_0)U(x_0,x_{m-1})\right)=\nn\\
~&=&\int_{S_{n}\times C_{m-3}(x_0,x_1,x_2)}
\int_{{\cal M}_\beta}{\left[\wedge_{j=n}^1{V^{(j)}_2}\right]}
\wedge tr
\wedge_{\alpha=m-1}^3{\left[W^{(\alpha)}_1{\cal U}(x_\alpha,x_{\alpha-1})\right]}\wedge\\ 
& &W^{(2)}_0(x_2){\cal U}(x_2,x_1)W^{(1)}_0(x_1){\cal U}(x_1,x_0)W^{(0)}_0(x_0)
{\cal U}(x_0,x_{m-1})~~,\nn
\eea}\noindent where the path integral is performed over the instanton sector 
$\beta$.   In  the  first  expression,  the  integral  over  boundary
insertions  is  performed over  the set  $C_{m-1}(x_0)$ of  
configurations
$(x_1..x_{m-1})$  such  that  $x_0,x_1,x_2...x_{m-1}$  are  distinct 
and cyclically
ordered on $\partial D$, and over the set 
$S_{n-1}(z_0)$ of bulk insertions points $z_1..z_n\in Int D$ such that the 
points $z_0..z_n$ are all distinct.   
A similar
convention is used in the second formula.

We  now  proceed to  give  a proof  of  these  equations.  Instead  of
adapting  the superspace  arguments of  \cite{DVV}, we  give  a direct
analysis in the universal  instanton formalism.  This has the advantage 
that it does not assume that the string background satisfies the 
string equations of motion. 

Recovering the nonlinear sigma  model formulation requires that we fix
the  $SL(2,\R)$  gauge  in   the  string  amplitude  $J_\beta$.  Since
(\ref{loc_string})  gives the  latter  in terms  of  the moduli  space
${\cal M}(\beta)$, it is not immediately clear what `fixing the gauge'
should mean.  The  solution is to express $J_\beta$  in terms of forms
defined on the  moduli space ${\tilde {\cal M}}(\beta)$,  on which the
symmetry group  $SL(2,\R)$ acts nontrivially. For this,  note that the
target space forms $v^{(j)}$ and  $w^{(\alpha)}$ can be pulled back to
${\tilde  {\cal  M}}(\beta)$ with  the  help  of  the evaluation maps.   
The resulting forms ${\tilde  V}^{(j)}$ and ${\tilde W}^{(\alpha)}$ on
${\tilde     {\cal    M}}(\beta)$     are     related    to     ${\hat
V}^{(j)}:=e_j^*(v^{(j)})$                  and                  ${\hat
W}^{(\alpha)}=e_\alpha^*(w^{(\alpha)})$ through:
\bea
{\tilde        V}^{(j)}&=&p^*({\hat        V}^{(j)})~~\\       {\tilde
W}^{(\alpha)}&=&p^*({\hat W}^{(\alpha)})~~.
\eea
This  follows  from  (\ref{ep}). We see  that ${\tilde  V}$ and
${\tilde W}$  are basic forms on  the fibration  
${\tilde {\cal M}}(\beta)\rightarrow {\cal M}(\beta)$
(i.e. pull-backs of  forms defined  over its  base).  
Given  a  section  $s$  of this  fibration,  the 
property  $p\circ s=id_{{\cal  M}(\beta)}$  implies that  $s^*({\tilde
V}^{(j)})={\hat   V}^{(j)}$   and  $s^*({\tilde   W}^{(\alpha)})={\hat
V}^{(\alpha)}$. Hence we can reformulate $J_\beta$ as follows:
\bea
\label{nlsm_string1}
J_\beta&=&\int_{{\tilde                                             {\cal
M}}(\beta)}{\left[\wedge_{j=n}^1{s^*({\tilde V}^{(j)})}\right]
\wedge tr\left[
\wedge_{\alpha=m-1}^0
{\left(s^*({\tilde W}^{(\alpha)}){\cal U}_{\alpha,\alpha-1}\right)}\right]}=\nn\\
&=&\int_{Im s}{\left[\wedge_{j=n}^1{{\tilde V}^{(j)}}\right]
\wedge tr\left[
\wedge_{\alpha=m-1}^0
{\left({\tilde W}^{(\alpha)}{\tilde {\cal U}}_{\alpha,\alpha-1}\right)}\right]}~~.
\eea

To  understand  the   precise  relation  between  ${\tilde  V}^{(j)}$,
${\tilde W}^{(\alpha)}$ and  $V^{(j)},W^{(\alpha)}$, consider the maps
$p_j:{\tilde {\cal  M}}(\beta)\rightarrow D\times {\cal  M}_\beta$ and
$p_\alpha:{\tilde {\cal M}}(\beta)\rightarrow  \partial D \times {\cal
M}_\beta$ defined through:
\bea
p_j(z,x,\phi)&:=&(z_j,\phi)~~\\
p_\alpha(z,x,\phi)&:=&(x_\alpha,\phi)~~.
\eea
It is easy to see that one has:
\bea
\Phi\circ p_j&=&{\tilde e}_j (=e_j\circ p)~~\\
\Phi_\partial\circ p_\alpha&=&{\tilde e}_\alpha(=e_\alpha\circ p)~~.
\eea
It follows that:
\bea
\label{pullbacks}
{\tilde   V}^{(j)}&=&p_j^*(V^{(j)})=p^*({\hat   V}^{(j)})~~\\  {\tilde
W}^{(\alpha)}&=&p_\alpha^*(W^{(\alpha)})=p^*({\hat W}^{(\alpha)})~~.
\eea
The relation between the various maps is summarized in the commutative
diagrams below.

\iffigs
\hskip 1.0 in
\begin{center} 
\scalebox{0.6}{\begin{picture}(0,0)%
\epsfbox{diagram.pstex}%
\end{picture}%
\setlength{\unitlength}{4144sp}%
\begingroup\makeatletter\ifx\SetFigFont\undefined%
\gdef\SetFigFont#1#2#3#4#5{%
  \reset@font\fontsize{#1}{#2pt}%
  \fontfamily{#3}\fontseries{#4}\fontshape{#5}%
  \selectfont}%
\fi\endgroup%
\begin{picture}(5490,7329)(1621,-6793)
\put(1936, 74){\makebox(0,0)[lb]{\smash{\SetFigFont{17}{20.4}{\familydefault}{\mddefault}{\updefault}${\tilde {\cal M}}^n_m(\beta)$}}}
\put(1981,-6496){\makebox(0,0)[lb]{\smash{\SetFigFont{17}{20.4}{\familydefault}{\mddefault}{\updefault}${\cal M}^n_m(\beta)$}}}
\put(1666,-5236){\makebox(0,0)[lb]{\smash{\SetFigFont{17}{20.4}{\familydefault}{\mddefault}{\updefault}$s$}}}
\put(2791,-5146){\makebox(0,0)[lb]{\smash{\SetFigFont{17}{20.4}{\familydefault}{\mddefault}{\updefault}$p$}}}
\put(4546,-3571){\makebox(0,0)[lb]{\smash{\SetFigFont{17}{20.4}{\familydefault}{\mddefault}{\updefault}$p_j$}}}
\put(4636,-6721){\makebox(0,0)[lb]{\smash{\SetFigFont{17}{20.4}{\familydefault}{\mddefault}{\updefault}$e_j$}}}
\put(4996,-5101){\makebox(0,0)[lb]{\smash{\SetFigFont{17}{20.4}{\familydefault}{\mddefault}{\updefault}${\tilde e}_j$}}}
\put(1936,-2581){\makebox(0,0)[lb]{\smash{\SetFigFont{17}{20.4}{\familydefault}{\mddefault}{\updefault}${\cal M}^n_m(\beta)$}}}
\put(7066,-1186){\makebox(0,0)[lb]{\smash{\SetFigFont{17}{20.4}{\familydefault}{\mddefault}{\updefault}$\Phi_\partial$}}}
\put(4951,-1186){\makebox(0,0)[lb]{\smash{\SetFigFont{17}{20.4}{\familydefault}{\mddefault}{\updefault}${\tilde e}_\alpha$}}}
\put(6706,-2581){\makebox(0,0)[lb]{\smash{\SetFigFont{17}{20.4}{\familydefault}{\mddefault}{\updefault}$L$}}}
\put(1621,-1321){\makebox(0,0)[lb]{\smash{\SetFigFont{17}{20.4}{\familydefault}{\mddefault}{\updefault}$s$}}}
\put(2746,-1231){\makebox(0,0)[lb]{\smash{\SetFigFont{17}{20.4}{\familydefault}{\mddefault}{\updefault}$p$}}}
\put(6436, 29){\makebox(0,0)[lb]{\smash{\SetFigFont{17}{20.4}{\familydefault}{\mddefault}{\updefault}$\partial D\times {\cal M}_\beta$}}}
\put(4501,344){\makebox(0,0)[lb]{\smash{\SetFigFont{17}{20.4}{\familydefault}{\mddefault}{\updefault}$p_\alpha$}}}
\put(4591,-2806){\makebox(0,0)[lb]{\smash{\SetFigFont{17}{20.4}{\familydefault}{\mddefault}{\updefault}$e_\alpha$}}}
\put(2026,-3796){\makebox(0,0)[lb]{\smash{\SetFigFont{17}{20.4}{\familydefault}{\mddefault}{\updefault}${\tilde {\cal M}}^n_m(\beta)$}}}
\put(7111,-5101){\makebox(0,0)[lb]{\smash{\SetFigFont{17}{20.4}{\familydefault}{\mddefault}{\updefault}$\Phi$}}}
\put(6481,-3886){\makebox(0,0)[lb]{\smash{\SetFigFont{17}{20.4}{\familydefault}{\mddefault}{\updefault}$D\times {\cal M}_\beta$}}}
\put(6751,-6496){\makebox(0,0)[lb]{\smash{\SetFigFont{17}{20.4}{\familydefault}{\mddefault}{\updefault}$X$}}}
\end{picture}
}
\end{center}
\begin{center} 
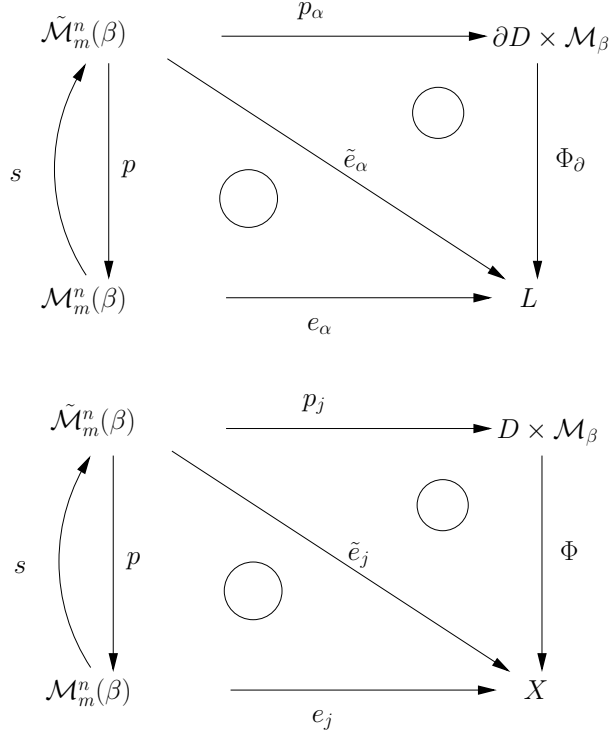
Figure 4. {\footnotesize Projection and evaluation maps.}
\end{center}
\fi

The  section $s$  implements  a  choice of  gauge  for the  $SL(2,\R)$
symmetry. To recover the nonlinear sigma model description, one simply
picks a gauge in which the positions of some insertion points are fixed
\footnote{Complete gauge-fixing via this procedure is not possible for string correlators
containing a low number of insertions. Such correlators 
are exceptional and are discussed in the next subsection.}.  Let us show
this explicitly for the case $m\ge  3$.  In this situation, we can fix
the  positions of  the  first  3 boundary  insertions  to some  values
$x_0,x_1,x_2$  by choosing  the  following section  of ${\tilde  {\cal
M}(\beta)}\rightarrow      {\cal     M}(\beta)$:     
{\footnotesize\be
s_{x_0,x_1,x_2}(q)=(z_1^q(z_0,x_1,x_2)...z_n^q(x_0,x_1,x_2),
x_0,x_1,x_2,x_3^q(x_0,x_1,x_2)...x_{m-1}^q(x_0,x_1,x_2),
\phi^q_{x_0,x_1,x_2})~~.
\ee}\noindent Here $q$ is a point in ${\cal M}(\beta)$, i.e. an orbit of the 
$SL(2,\R)$ action on 
${\tilde {\cal M}}(\beta)$. In the right hand side, we have picked the
unique  point $(z,x,\phi)$  on this  orbit with  the property  the the
first  3 components  of $x$  coincide with  $x_0,x_1$ and  $x_2$. This
determines  the  remaining  components  $x_3...x_{m-1}$,  as  well  as
$z_1..z_n$ in  terms of the orbit  $q$ and the  fixed points $x_0,x_1$
and    $x_2$.    It    also    determines    a    special    instanton
$\phi^q_{x_0,x_1,x_2}$.    Due    to   the   form    of   the   action
(\ref{SL_action}), $\phi^q_{x_0,x_1,x_2}$ will cover the entire moduli
space of  instantons ${\cal  M}_\beta$ precisely once,  as we  let the
orbit  $q$   cover  ${\cal  M}(\beta)$.  Notice  that   the  image  of
$s_{x_0,x_1,x_2}$  coincides   with  the  domain   of  integration  of
(\ref{string2}):
\be
Im s_{x_0,x_1,x_2}=S_n\times C_{m-3}(x_0,x_1,x_2)
\times {\cal M}_\beta~~.
\ee
Combining  this   observation  with  properties   (\ref{pullbacks}) and 
(\ref{nlsm_string1}),  it
immediately follows that (\ref{loc_string}) implies (\ref{string2}). \
A similar argument applies for (\ref{string1}).

\subsection{Exceptional string amplitudes and instanton corrections to the 
localized BRST operator}

The disk string amplitudes containing one or two boundary insertions
and no bulk insertion or one bulk insertion and no boundary insertions 
are special from a few points of view. These string correlators cannot
be expressed as sigma model amplitudes of descendants, due to the fact that 
fixing the positions of operator insertions 
does not suffice to fix the $SL(2,\R)$ gauge. In particular, one cannot use 
the sigma model non-renormalization results in order to conclude that the 
{\em string} amplitudes $\langle \langle {\cal O}_v\rangle\rangle$ and 
$\langle \langle {\cal O}_w\rangle \rangle$, $\langle \langle {\cal O}_{w_1}
{\cal O}_{w_2}\rangle \rangle$ do not receive instanton corrections.  
As we discuss in \cite{sft}, the presence of such amplitudes is related to the 
fact that the pair $(L,A)$ gives a correct description of the topological 
D-brane only in the large radius limit. Since we express 
string amplitudes in terms of the large radius data $(L,E)$, we are building 
a topological string theory around a large radius vacuum, which generally 
differs from the true vacuum. The resulting string amplitudes can be brought to
the standard form upon performing a shift from the large radius vacuum 
to the true vacuum \cite{sft}.

Let us discuss the three exceptional classes of string amplitudes. 
We start with the {\bf two-point boundary amplitude}, which has the 
instanton expression:
\be
\label{2pt}
\langle \langle {\cal O}_{w_1}{\cal O}_{w_2}\rangle \rangle=
\sum_{\beta}{\langle \langle {\cal O}_{w_1}{\cal O}_{w_2}\rangle \rangle_\beta
e^{-S_\beta}}~~. 
\ee

The coefficients $\langle \langle {\cal O}_{w_1}{\cal O}_{w_2}\rangle\rangle_\beta$ 
in this expansion are defined through integrals over ${\cal M}^0_2(\beta)$ as in 
(\ref{loc_string}). One can fix the position of the two 
boundary insertion points at $x_1$ and $x_2$ as before, but this does not suffice to 
define a section of the fibration ${\tilde {\cal M}}^0_2(\beta)\rightarrow {\cal M}^0_2(\beta)$. 
Instead, this procedure gives a presentation of ${\cal M}^0_2(\beta)$ as a quotient 
of ${\cal M}_\beta$ through the subgroup $G\approx \R$ of $SL(2,\R)$ which fixes the points 
$x_1$ and $x_2$. Hence one can express $\langle \langle {\cal O}_{w_1}{\cal O}_{w_2}
\rangle\rangle_\beta$ as an integral over ${\cal M}_\beta/\R$. This differs from the 
the nonlinear sigma model amplitude $\langle {\cal O}_{w_1}(x_1){\cal O}_{w_2}(x_2)
\rangle_\beta$, which is expressed as an integral over ${\cal M}_\beta$. In particular, 
the latter depends on the insertion points $x_1$ and $x_2$ while the former is independent 
of this choice.

The trivial instanton term in the expansion (\ref{2pt}) is exceptional, since the 
action of $SL(2,\R)$ on the moduli space ${\tilde {\cal M}}^0_2(0)=C^2\times L$ 
is not fixed-point free. To recover the correct value of 
$\langle \langle {\cal O}_{w_1}{\cal O}_{w_2}\rangle \rangle^{(0)}$,
one must perform a direct analysis of localization starting with 
the Polyakov path integral
\footnote{In this approach, the appearance of the differential 
$d_A$ in (\ref{2pt_0}) is related to the existence of fixed points for 
the $SL(2,\R)$ action on ${\tilde {\cal M}}^0_2(0)$.}. 
Alternatively, one can use the results of 
\cite{Witten_CS}, which studied this localization (in the Hamiltonian 
approach) for the trivial instanton sector
\footnote{Most of the analysis of \cite{Witten_CS} 
is restricted to the case when $X=T^*M$ for some manifold $M$. This assumption 
is made in \cite{Witten_CS} with the sole purpose of eliminating instanton 
corrections. Indeed, all contributions from nontrivial 
instanton sectors vanish for such manifolds, 
due to a vanishing theorem proved in \cite{Witten_CS}. However, the entire 
analysis  of \cite{Witten_CS} remains correct in the trivial instanton sector 
($\beta=0$) or, equivalently, in the large radius limit for an arbitrary 
Calabi-Yau manifold $X$.}. It follows from the results of 
\cite{Witten_CS} that the relevant two-point amplitude in this sector 
must be:
\be
\label{2pt_0}
\langle \langle {\cal O}_{w_1}{\cal O}_{w_2}\rangle \rangle^{(0)}=
\frac{1}{r}\int_{L}{tr(w_1\wedge d_A w_2)}~~.
\ee
(this gives the quadratic term of the large radius string field theory action 
of \cite{Witten_CS}). Note that (\ref{2pt_0}) has no direct relation to the 
trivial instanton sector contribution to the sigma-model boundary two-point amplitude. 
In particular, the latter is non-vanishing if $rank w_1+rank w_2=d$, while the selection rule 
for (\ref{2pt_0}) is $rank w_1+rank w_2=d-1$.

The higher instanton corrections to (\ref{2pt}) can be encoded by writing:
\be
\langle \langle {\cal O}_{w_1}{\cal O}_{w_2}\rangle \rangle=\rho(w_1,Q_ow_2)=
\frac{1}{r}\int_{L}{tr\left[w_1\wedge Q_ow_2\right]}~~,
\ee
where $Q_o=\sum_{\beta}{Q_o(\beta)}$, with
$Q_o(0)=d$ and $Q_o(\beta):\Omega^*(L,E)\rightarrow \Omega^*(L,E)$ some 
degree $+1$ linear maps determined by:
\be
\langle \langle {\cal O}_{w_1}{\cal O}_{w_2}\rangle \rangle^{(\beta)}=
\rho(w_1,Q_o(\beta)w_2)=\frac{1}{r}\int_{L}{tr\left[w_1\wedge Q_o(\beta)w_2
\right]}~~.
\ee
The fact that the higher instanton contributions
$Q_o(\beta)$ $(\beta\neq 0)$ 
need not vanish was noticed for the first time in \cite{Fukaya}. 
The operator $Q_o$ can be viewed as an instanton-corrected expression for 
the BRST charge of the model. To prevent misunderstanding
\footnote{I thank R.~Roiban for pointing out this possible source of 
confusion.}, let me 
make a few observations on the physical meaning of this interpretation. 
In \cite{Witten_CS}, it was argued that the BRST charge of the model 
can be represented (in the large radius limit) as the differential $d_A$. 
This follows from a localization 
argument in the {\em trivial} instanton sector $\beta=0$, and 
cannot be extended to nontrivial sectors. The fact that 
$Q_o$ receives worldsheet instanton corrections refers to the 
realization of the BRST symmetry on the space $\Omega^*((End(E))$,
i.e. it is a statement about the nature of localization. 
As we shall see in \cite{sft}, the presence of such corrections modifies the 
string field action. It is important, however, 
to realize that such effects modify the {\em classical} BV action of the string 
field theory, and do not represent a quantum (i.e. loop) 
effect in string theory per se (even though
they do represent a quantum effect on the {\em worldsheet}). 
We do {\em not} claim that the BRST operator receives corrections (and, in 
fact, may become `anomalous') due to the quantum dynamics governed by 
the string field action\footnote{Such an effect may also occur, but I am not 
able to test that at this point.}. 
Rather, we are saying that the {\em tree level} 
string field action of \cite{Witten_CS} is incomplete due to worldsheet effects 
which modify the localization formula for the BRST charge.  

In fact, the localized BRST operator is given by a `shifted' form $Q'_o$ of 
$Q_o$. This further modification of $Q_o$ 
is induced by the necessity of performing a 
shift of the string background, as I discuss in 
more detail in \cite{sft}. Indeed, the 
operator $Q_o$ does not generally square to zero, as already noticed for the 
singly-wrapped case by Fukaya and collaborators \cite{Fukaya, Fukaya2}. It is the 
modification $Q_o\rightarrow Q'_o$ which assures that the shifted operator $Q'_o$
is nilpotent. 

Finally, we consider the {\bf one-point string  amplitudes} 
$\langle \langle {\cal O}_v\rangle \rangle$ and $\langle \langle {\cal O}_w
\rangle \rangle$. These can be written:
\bea
\langle \langle {\cal O}_v\rangle \rangle &=&(v,p^0_0)~~,\\
\langle \langle {\cal O}_w\rangle \rangle &=&\rho(w,q^0_0)~~,
\eea
for some elements $p^0_0\in \Omega^{3}(X)$ and $q^0_0\in \Omega^2(L,End(E))$.
Here $(,)$ is a degree $2d-1$ bilinear form which is induced from the 
bulk topological metric when restricting to the semirelative complex of the 
closed string state space \cite{sft}. Since this requires a slightly lengthy analysis
(to be given in \cite{sft}), 
I will concentrate on the amplitude $\langle \langle {\cal O}_w\rangle \rangle$.

As before, one has instanton expansions:
\bea
\label{q_exp}
\langle \langle {\cal O}_w\rangle \rangle &=&
\sum_{\beta}{\langle \langle {\cal O}_w\rangle \rangle_\beta e^{-S_\beta}}~~\\
q^0_0 &=&\sum_{\beta}{q^0_0(\beta)e^{-S_\beta}}~~,
\eea
with $q^0_0(\beta)\in \Omega^2(L,End(E))$ defined through:
\be
\langle \langle {\cal O}_w\rangle \rangle_\beta=
\frac{1}{r}\int_{{\cal M}^0_1(\beta)}{tr(e_0^*(w){\cal U}_{0,0})}
=\rho(w,q^0_0(\beta))~~,
\ee
where ${\cal U}_{0,0}\in End({\cal E}_0)$ is induced by 
the holonomy operator ${\tilde {\cal U}}_{0,0}\in End({\tilde {\cal E}}_o)$ 
associated with a curve which winds once around the cylinder 
${\partial D}\times {\cal M}_\beta$. Upon fixing the 
boundary insertion point to $x_o$, the space ${\cal M}^0_1(\beta)$ can be presented 
as the quotient of ${\cal M}_\beta$ through the two-dimensional stabilizer $G$ of $x_o$ in 
$SL(2,\R)$. In the conformal case, this gives a moduli space of dimension 
$dim {\cal M}^0_2(\beta)= d-2=d+1-3$, which implies that $q^0_0(\beta)$ is a 
form of rank $2$. For $\beta=0$, the moduli space ${\cal M}^0_1(0)$ 
coincides with ${\cal M}_0=L$, and no rank two element of $\Omega^*(L,End(E))$ can 
be produced in this manner\footnote{Inserting the differential $d_A$ 
is not allowed, since this would lead to a form of degree $3$. That $q^0_0(0)$ must vanish
also follows from the general structure of open string field theory \cite{Gaberdiel} and 
the fact that our background does satisfy the string equations of 
motion at large radius (see below).} 
Therefore, one must take $q^0_0(0)=0$. In particular, the 
sums in (\ref{q_exp}) can be taken to run over nontrivial homotopy classes $\beta\neq 0$. 
It follows that the large radius limit of $q^0_0$ vanishes, and the presence of 
a nonzero $q^0_0$  is purely an instanton effect. 

The fact that $q^0_0$ need not vanish was noticed for the first time in 
\cite{Fukaya}. As I discuss in more detail in \cite{sft}, the presence of 
this string product is a signal that the string background receives quantum 
corrections; according to (a slight generalization of) \cite{Gaberdiel}, 
this means that we are building a string field theory around 
the wrong vacuum.
Note that the product $q^0_0$ need not be zero even if the 
cycle $L$ is special Lagrangian. In the special Lagrangian case, this is 
responsible for the obstructions to D-brane deformations observed in 
\cite{Douglas_quintic,Kachru}. This zeroth order product can be eliminated by 
shifting the string background \cite{Fukaya, Fukaya2, sft}. As for $p^0_0$, it can 
also be eliminated by shifting the string background \cite{sft}. 

\subsection{Homological formulation for the boundary sector 
of a singly-wrapped D-brane}

When  $E$ is  a line  bundle on  $L$, the boundary nonlinear sigma  model and
string  amplitudes admit  a  homological formulation  similar to  that
familiar from the  closed string case.  In this  situation, the bundle
$End(E)\approx  E^*\otimes E$  is trivial,  and  hence $End(E)$-valued
forms on $L$ become usual complex-valued forms. The connection induced
by $A$  on $End(E)$ is a  flat connection on ${\cal  O}_L$. While $End(E)$ is
trivial, the  parallel transport of  $A$ must still be taken
into account.

The  homological  form  of  the boundary amplitudes  results  by  
considering forms $w_\alpha$ with delta-function support
on submanifolds $K_\alpha$ of $L$, having 
codimension $l_\alpha$ (that is, $w_\alpha$ are currents 
associated to such submanifolds). Then the
pull-backs  of $w_\alpha$  by the  various evaluation
maps used  in the universal  instanton formalism have delta-function 
support on codimension $l_\alpha$ cycles $e_{x_\alpha}^*(K_\alpha)$ 
of ${\cal M}_\beta$ and $e_\alpha^*(K_\alpha)$ 
of ${\cal M}^0_m(\beta)$. Integration of  the wedge
product  over the  moduli space  amounts to intersection of the
pulled-back  cycles.  The product is zero unless the 
common intersection of the pulled-back cycles is a discrete set of points. 
It follows that the localization formulae reduce to 
finite  sums over these intersection points. At each intersection 
point $\phi$, the parallel 
transport operators ${\cal U}$ can be composed, giving the  
holonomy in the fiber of the pulled-back line bundle ${\cal E}$ 
above that point, which coincides 
with the  holonomy 
$W_\beta=Pe^{-\int_{\partial  D}{\phi^*(A)}}$ of 
of  $-\phi^*(A)$  around  $\partial D$. 
This  quantity depends only on the relative
homotopy class $\beta$, since the closed curves $\phi(\partial D)$ and
$\phi'(\partial D)$ are  homotopic to each other {\em  in L} if $\phi$
and $\phi'$  belong to $\beta$. 

The net effect is to produce the intersection form
of  the pulled-back  cycles, modulo  the holonomy factor $W_\beta$. 
For the boundary sigma model amplitude, the end result is:
\be
\label{homological1}
I_\beta(K_{m-1}..K_0)=\#(\cap_{\alpha=m-1}^0{
(e_{x_\alpha})^* (K_\alpha)})_{{\cal M}_{\beta}}W_\beta~~,
\ee
while for the boundary 
string amplitude one obtains:
\be
J_\beta(K_{m-1}..K_0)=\#(\cap_{\alpha=m-1}^0{e_{\alpha}^*
(K_\alpha)})_{{\cal M}^0_m(\beta)}W_\beta~~.
\ee
This allows us to extract
a  homological formulation of  string products.  Indeed, the 
boundary topological  metric corresponds to the  intersection form on
$L$:
\be
{\check \rho}(K_\alpha,K_\beta)=\#(K_\alpha\cap
K_\beta)_L=\int_{L} {v^{(\alpha)}\wedge v^{(\beta)}}~~.
\ee
Hence we can write, for example:
\be
J_\beta(K_{m}..K_0)=
{\check \rho}(K_{m},~{\check r}_{m}(\beta)(K_{m-1}..K_{0}))~ W_\beta~~,
\ee
where          ${\check r}_{m}(\beta)(K_{m-1}..K_0)
=(e_{m})_*(
e_{m-1}^*(K_{m-1})\cap ...\cap e_{0}^*(K_{0}))$. This gives the dual 
formulation of the boundary string products:
\be
\label{string_product}
{\check r}_m =
\sum_{\beta}{{\check r}_m(\beta) W_\beta e^{-S_\beta}}~~.
\ee 
${\check  r}_m$  is a  linear  map  from 
$C_*(L)^{\otimes m}$  to $C_*(L)$. 
Relation  (\ref{string_product})  recovers  the
products introduced by Fukaya and collaborators \cite{Fukaya, Fukaya2}.

Let me end with a few technical remarks for the mathematically savvy reader. 
The `off-shell Poincare duality' argument employed above is quite non-rigorous, 
due to technical difficulties involved in defining the intersection theory of 
chains. In particular, it is not obvious how the argument works for 
the case when two of the chains $K_\alpha$ coincide. This is related to 
a host of technical issues discussed in detail in \cite{Fukaya2}, to which 
I refer the interested reader. Unfortunately, similar problems are encountered 
in the cohomological formulation (for the singly or multiply-wrapped 
case), when one attempts to define string products 
off-shell via dualization with respect to the topological metrics. It remains 
a difficult problem to provide a rigorous analysis in this 
framework. 

Finally, let me make a few remarks on the possibility of a homological approach for 
the multiply-wrapped case. It is intuitively clear that recovering a homological approach 
requires the use of `bundle-valued currents'. This is to say that the homological 
formulation of the boundary state space must include the Chan-Paton degrees of freedom
described by the local system (flat bundle) $E$. One could try to achieve this by 
considering chains $K$ in $L$ endowed with extra data represented by a section of 
$End(E)$ above $K$, but it is not clear to what extent such a description is 
equivalent to the cohomological approach proposed in this paper.

\appendix

\section{Localized form of the BRST operator in the large radius limit}

This appendix re-derives the BRST invariance 
condition for the boundary observables in the `Lagrangian' framework 
of \cite{Witten_NLSM, Witten_mirror}. The result is well-known from 
\cite{Witten_CS}, but there seems to be some confusion on this point 
in the literature, so I include a detailed derivation for completeness.  

To  identify   the  boundary  topological  observables,   it  is  most
convenient     to    follow     the    `Lagrangian'     approach    of
\cite{Witten_NLSM,Witten_mirror}. 
Let  $Q_o$ denote  the boundary  BRST generator,  which  satisfies $\left[
Q_o,\phi\right]=-\chi$ and $\{Q_o,\chi\}=0$. We  want to express the constraint
$\{Q_o,{\cal  O}_{w}\}=0$  as  a  condition on  the  bundle-valued  form
$w$. To understand the structure of  (\ref{obs1}), let us first look 
at its form in the 
vicinity of a  given  point   $x\in  {\partial D}$.   Choosing   suitable  open
neighborhoods $U$ of $\phi(x)$  and $V\subset \phi^{-1}(U)$ of $x$, we
can    trivialize    $E|_{U}$    with    the    help    of    a    set
$\{s_\alpha\}_{\alpha=1..r}$ of linearly independent sections
\footnote{Linearly independent at any point of $U$, i.e. a local frame for $E$ 
above     $U$.}.     We    also     consider     the    dual     frame
$\{s^*_\alpha\}_{\alpha=1..r}$   of    $E^*|_{U}$,   which   satisfies
$s^*_\alpha(s_\beta)=\delta_{\alpha,\beta}$.   Finally,  we consider  a
coordinate  chart  of  $L$  on  $U$,  which  defines  tangent  vectors
$\{\partial_i\}_{i=1..d}$ to $U$.
Let  $\nabla$   be  the  D-brane   connection  on  $E$  and   $A$  its
matrix-valued coefficient one-form in the frame $s$:
\be
d_\nabla(s_\alpha)=A_{\beta\alpha}s_\beta~~.
\ee
We will  also need the dual  connection $\nabla^*$ on  $E^*$, which is
defined through the condition:
\be
X[\eta(s)]=[\nabla^*_X(\eta)](s)+\eta(\nabla_X(s))~~,
\ee 
for any local  sections $\eta$ of $E^*$ and $s$  of $E$, and any 
vector field $X$ on $U$.  Its one-form
coefficient matrix in the frame $s^*$ is given by:
\be
d_\nabla^*(s^*_\alpha)=A^*_{\beta\alpha}s^*_\beta=-A_{\alpha\beta}s^*_\beta~~,
\ee
i.e. $A^*=-A^t$.

With these preparations, we can write the bundle-valued form $w$ as:
\be
w=w^{\beta\alpha}s^*_\beta\otimes s_\alpha~~,
\ee
and the observable ${\cal O}_w$:
\be
{\cal O}_w(x)=w^{\beta\alpha}_{i_1..i_k}(\phi(x))s^*_\beta(\phi(x))
\otimes s_\alpha(\phi(x))\chi^{i_1}(x)...\chi^{i_k}(x)~~, 
\ee
where    $w^{\beta\alpha}_{i_1..i_k}=w^{\beta\alpha}(\partial_{i_1},...
,\partial_{i_k})$.

Before attempting  to compute  the BRST variation  of the object (\ref{obs1}), 
we have to understand  if we  have actually  provided a
complete definition of what its  BRST variation should be (as we shall
see in a moment, the answer is negative). A complete definition of the
model  requires that  we  define how  the  operation $Q$  acts on  all
`allowed'  configurations.  The later  are  functionals  of the  basic
worldsheet  fields $\phi,\chi,\lambda$.  The standard  approach, followed
for  example in  \cite{Witten_NLSM, Witten_mirror,  Witten_CS},  is to
specify the Q-variation of these basic fields and to define the action
of  $Q$   on  an  arbitrary   functional  $F[\phi,\chi,\lambda]$  through
functional differentiation:
\be
(\delta_QF)[\phi,\chi,\lambda]=
\int_{\partial D}{\left({\frac{\delta        F}{\delta
\phi(x)}(\delta_Q\phi)(x)+
\frac{\delta F}{\delta \chi(x)}(\delta_Q\chi)(x)+
\frac{\delta F}{\delta \lambda(x)}(\delta_Q\lambda)(x)}\right)}~~.
\ee
However, this assumes that are given a proper definition of the 
objects $\frac{\delta  F}{\delta \phi}(x)$. 
Such a definition is not immediately obvious, since we want
to  functionally differentiate a geometrically nontrivial 
object $F$. Indeed,  in our case $F[\phi,\chi]={\cal O}_w$
is a  bundle-valued form on ${\partial D}$. Its formal Q-variation is given by:
\be
\delta_Q{\cal O}_w=i\xi
\int_{{\partial D}}{\frac{\delta {\cal O}_w}{\delta \phi(x)}\chi(x)}~~.
\ee
To define the functional variation  in the integrand, one must specify
how     to     compare     the     values    of     the     functional
$f[\phi]=s^*_\beta\circ\phi\otimes    s_\alpha\circ\phi$    for    two
different choices  $\phi_1$ and $\phi_2$ of $\phi$.  In particular, we
must know how to compare  the values of $s^*_\beta\otimes s_\alpha$ at
two  {\em different} points  $\phi_1(x)$ and  $\phi_2(x)$ on  $L$. The
standard procedure for  doing this is to consider  a connection on the
bundle  $End(E)$  and compare  the  two  values  through the  parallel
transport it defines. Hence a  complete {\em definition} of the action
of  $Q$   in  the  boundary  sector   requires  the  use  of a
connection. There is only one natural candidate arising 
from our  model's data, namely  the connection induced by  $\nabla$ on
$End(E)$. Hence we {\em define}:
\bea
\delta_Q s_\alpha&=&i\xi(d_\nabla s_\alpha)_x(\chi(x))=
i\xi\nabla_is_\alpha(x)\chi^i(x)~~\\
\delta_Q s^*_\alpha&=&
i\xi(d_{\nabla^*}s^*_\alpha)_x(\chi(x))=i\xi\nabla^*_is^*_\alpha(x)\chi^i(x)~~.
\eea 
This allows us to compute  the variation of $s^*_\alpha\circ \phi
\otimes s_\beta\circ\phi$: {\footnotesize
\bea
\delta_Q 
(s^*_\alpha\circ             \phi\otimes            s_\beta\circ\phi)=
i\xi[(\nabla_is^*_\alpha)(\phi(x))\otimes                s_\beta(\phi(x))+
s_\alpha(\phi(x))\otimes(\nabla_is_\beta)(\phi(x))]\chi^i(x)\\        =
i\xi[-A_{\alpha\alpha'~i}(\phi(x))\delta_{\beta'\beta}+\delta_{\alpha,\alpha'}
A_{\beta'\beta,i}(\phi(x))          ]\chi^i(x)s^*_\beta(\phi(x))\otimes
s_\alpha(\phi(x))~~.
\eea}
One also has:
\be
\delta_Q(w^{\beta\alpha}_{i_1..i_k}\circ \phi)(x) =
i\xi(\partial_iw^{\beta\alpha}_{i_1..i_k})(\phi(x))\chi^i(x)~~.
\ee
Combining these two results we obtain:
\be
\{Q_o,{\cal O}_w\}=-{\cal O}_{d_Aw}~~,
\ee
where  $d_A$  is  the  covariant differential induced by  $A$  on  $End(E)\approx
E^*\otimes E$:
\be
d_Aw=dw+[A,w]~~.
\ee

\section{The Maslow index}

In this appendix  I discuss the quantity $\mu(\beta)$  which appears in
the virtual dimension of disk instanton moduli spaces \cite{Fukaya}.
Let us start by fixing a  Lagrangian cycle $L$ in $X$.  Consider a map
$\phi:D\rightarrow  X$   from  the  unit   disk  to  $X$,   such  that
$s:=\phi(\partial  D)\subset  L$.  We  let $\beta$  denote  its  relative
homotopy  class  in  $X$  with  respect to  $L$.  Since  $D$  is
contractible,  one  can globally  trivialize  the  pulled back 
tangent bundle $\phi^*(TX)$  on $D$.   This   result  also   holds
symplectically,  i.e. the trivialization  can be  chosen such  that it
takes the symplectic form $\omega_x (x\in D)$ on each fiber into
the  standard  symplectic  form  $\omega_d$  on  $\C^d=\R^{2d}$.  Here
$\omega$  is the pulled-back Kahler  form of  $X$.  
Since  $L$ is  Lagrangian, the
tangent  spaces $\phi^*(TL)_x$  for $x  \in \partial  D$  are Lagrangian
subspaces  of  $(\phi^*(TX)_x,\omega_x)$.  The  symplectic  trivialization  of
$\phi^*(TX)$  allows us to  view the  collection of  such subspaces  as a
curve $\Gamma$ in the  Lagrangian Grassmannian $Lagr_d$. The latter is
defined  as the  set of  all  Lagrangian subspaces  of the  symplectic
vector  space $(\R^{2d},\omega_d)$\footnote{A Lagrangian  subspace $V$
of a  symplectic vector  space is a  subspace whose dimension  is half
that  of  the  ambient space  and  such  the  the restriction  of  the
symplectic form to that subspace  is zero.}.  This can be formalized 
by considering the {\em Gauss map} $G:L\rightarrow Lagr(TX)$, which associates
to each point $u$ of $L$ the tangent bundle $T_uL\subset T_uX$. By using the 
 trivialization of $\phi^*(X)$, this induces a map $g:\partial D\rightarrow 
Lagr_d$. The curve $\Gamma$ is the image of 
$\partial D$ through $g$..

It is a classical result
that $\pi_1(Lagr_d)\approx  \Z$, with a generator which  we shall call
$\gamma$. Hence  the homotopy  class of the  curve $\Gamma$  inside of
$Lagr_d$ can be expressed as a multiple of this generator:
\be
\Gamma=\mu(\phi) \gamma~~.
\ee
The integer  $\mu$ is the {\em  Maslow index} of $\phi$. This
number is independent of  the trivialization chosen for $\phi^*(TX)$, since
a change  of this trivialization induces  a homotopy transformation  of 
$\Gamma$.   Moreover, the  Maslow  index  of $\phi$ depends  
only on  the
relative homotopy  class $\beta$, since two  maps $\phi_1$,
$\phi_2$ in the same  relative homotopy class induce homotopic
curves   $\Gamma_1,\Gamma_2$  in   $Lagr_d$.   Hence   we   can  write
$\mu(\phi)=\mu(\beta)$.

Intuitively,  the index  tells us  how  many times  the tangent  space
$T_uL$ `rotates' inside of $T_uX$  as $u$ moves once around the closed
curve $s=\phi(\partial  D)$.  
This  quantity appears in the index  theorem of \cite{Fukaya, Fukaya2} because the
boundary conditions for  $\chi$ along $\partial D$ require  it to be a
section  of  $\phi^*(TL)$.  Hence   the  `winding'  of  $T_uL$  ($u\in
\phi(\partial D)$) inside of $T_uX$ is related to the `winding' of the
fermionic section $\chi|_{\partial D}$  as one follows the boundary of
the disk.

\end{document}